\documentclass[prc,aps,twocolumn,showpacs,amssymb,superscriptaddress,fleqn]{revtex4-1}

\usepackage{amsmath}
\usepackage{color}
\usepackage[table]{xcolor}
\usepackage[english]{babel}
\usepackage{graphicx}
\usepackage{dcolumn}
\usepackage{mathtools}
\usepackage{relsize}
\usepackage{float}
\usepackage{braket}

\newcommand{\vlwk}{$V_{{\rm low}\mbox{-}k}$}
\newcommand{\thetaeff}{$\Theta_{\rm eff}$}

\newcommand{\zbb}{$0\nu\beta\beta$}
\newcommand{\dbb}{$2\nu\beta\beta$}
\newcommand{\heff}{$H_{\rm eff}$}
\newcommand{\heffs}{$H_{\rm eff}$'s}
\newcommand{\qbox}{$\hat{Q}$~box}
\newcommand{\tbox}{$\hat{\Theta}$~box}
\newcommand{\nme}{$M^{0\nu}$}
\newcommand{\nmes}{$M^{0\nu}$'s}
\newcommand{\nmed}{$M^{2\nu}$}
\newcommand{\nmeds}{$M^{2\nu}$'s}
\newcommand{\nmax}{$N_{\rm max}$}

\begin{document}

\title{The renormalization of the shell-model Gamow-Teller operator starting from effective field theory for nuclear systems}

\author{L. Coraggio}
\affiliation{Dipartimento di Matematica e Fisica, Universit\`a degli
  Studi della Campania ``Luigi Vanvitelli'', viale Abramo Lincoln 5 -
  I-81100 Caserta, Italy}
\affiliation{Istituto Nazionale di Fisica Nucleare, \\ 
Complesso Universitario di Monte  S. Angelo, Via Cintia - I-80126 Napoli, Italy}
\author{N. Itaco}
\affiliation{Dipartimento di Matematica e Fisica, Universit\`a degli
  Studi della Campania ``Luigi Vanvitelli'', viale Abramo Lincoln 5 -
  I-81100 Caserta, Italy}
\affiliation{Istituto Nazionale di Fisica Nucleare, \\ 
Complesso Universitario di Monte  S. Angelo, Via Cintia - I-80126 Napoli, Italy}
\author{G. De Gregorio}
\affiliation{Dipartimento di Matematica e Fisica, Universit\`a degli
  Studi della Campania ``Luigi Vanvitelli'', viale Abramo Lincoln 5 -
  I-81100 Caserta, Italy}
\affiliation{Istituto Nazionale di Fisica Nucleare, \\ 
Complesso Universitario di Monte  S. Angelo, Via Cintia - I-80126 Napoli, Italy}
\author{A. Gargano}
\affiliation{Istituto Nazionale di Fisica Nucleare, \\
Complesso Universitario di Monte  S. Angelo, Via Cintia - I-80126 Napoli, Italy}
\author{Z. H. Cheng}
\affiliation{School of Physics and Key State Laboratory of Nuclear
  Physics and Technology, \\
Peking University, Beijing 100871, China}
\author{Y. Z. Ma}
\affiliation{School of Physics and Key State Laboratory of Nuclear
  Physics and Technology, \\
Peking University, Beijing 100871, China}
\author{F. R. Xu}
\affiliation{School of Physics and Key State Laboratory of Nuclear
  Physics and Technology, \\
Peking University, Beijing 100871, China}
\author{M. Viviani}
\affiliation{Istituto Nazionale di Fisica Nucleare, \\
Largo Bruno Pontecorvo 3 - I-56127 Pisa, Italy}

\begin{abstract}
For the first time, we approach in this work the problem of the
renormalization of the Gamow-Teller decay operator for nuclear
shell-model calculations by way of many-body perturbation theory,
starting from a nuclear Hamiltonian and electroweak currents derived
consistently by way of the chiral perturbation theory.
These are the inputs we need to construct microscopically the
effective shell-model Hamiltonians and decay operators.
The goal is to assess the role of both electroweak currents and
many-body correlations as the origins of the well-known problem of the
quenching of the axial coupling constant $g_A$.
To this end, the calculation of observables related to the
Gamow-Teller transitions has been performed for several nuclear
systems outside the $^{40}$Ca and $^{56}$Ni closed cores and compared
with the available data.
\end{abstract}

\pacs{21.60.Cs, 21.30.Fe, 27.40.+z}

\maketitle
\section{Introduction}
\label{intro}
In recent years there has been a renewed interest to study the process
of $\beta$ decay of atomic nuclei in terms of nuclear structure
theoretical models
\cite{Pirinen15,Brown15,Simkovic18,Delion17,Suhonen17a,Coraggio17a,Simkovic18,Coello18,Coello19,Suhonen19,Coraggio19a,Deppisch20,Gambacurta20,Coraggio22a}.
In fact, this kind of investigation may provide insight about the
mechanism of neutrinoless double-$\beta$ decay (\zbb), and also an
important testing ground to validate the calculation of the nuclear
matrix element \nme~ of such a rare decay.
An important issue which most nuclear structure calculations have to
face is the overestimation of the Gamow-Teller (GT) transition rates,
and this defect is usually treated by quenching the axial coupling
constant $g_A$ by a factor $q<1$
\cite{Suhonen13,Ejiri83,Martinez-Pinedo96,Barea15,Suhonen19}.
The need to introduce effective values of $g_A$ in nuclear structure
calculations traces back to two main sources:
\begin{enumerate}
\item Nucleons are not pointlike particles, and their quark structure
  needs to be accounted for; namely the effects of meson-exchange
  currents (two-body electroweak currents) have to be considered
  \cite{Arima73}.
\item Apart from the ab initio approaches, all other nuclear models
  adopt a truncation of the full Hilbert space of the configurations
  of the nuclear wave functions into a reduced model space where a
  selected number of the degrees of freedom are retained. 
This operation is necessary to allow the diagonalization of the
nuclear Hamiltonian and, consequently, effective Hamiltonians and
decay operators must be introduced to account for the configurations
which have been neglected to construct the model-space nuclear wave
function \cite{Arima73}.
\end{enumerate}

The quenching of the axial coupling constant $g_A$ may have massive
impact on the estimate of the half-life of \zbb~ decay
$T^{0\nu}_{1/2}$.
Indeed, the latter is connected to the structure of the parent and
granddaughter nuclei by way of the nuclear matrix element \nme~
according to the following expression:
\begin{equation}
\left[ T^{0\nu}_{1/2}\right]^{-1} = G^{0\nu} \left| M^{0\nu} \right|^2
\left| f(m_i,U_{ei})\right|^2 ~,
\end{equation}
\noindent
$G^{0\nu}$ being the so-called phase-space factor (or kinematic
factor) \cite{Kotila12}, and  $f(m_i,U_{ei})$ accounting for the
adopted model of \zbb~ decay (light and/or heavy neutrino exchange,
etc.) by way of the neutrino masses $m_i$ and their mixing matrix
elements $U_{ei}$.
The explicit form of $f(m_i,U_{ei})$, within the mechanisms of
light-neutrino exchange, is $f(m_i,U_{ei})= g_A^2 \frac{ \langle
  m_{\nu}\rangle}{m_e}$, where $m_e$ is the electron mass, and
$\langle m _{\nu} \rangle = \sum_i (U_{ei})^2 m_i$ is the effective
neutrino mass.

The above expression evidences the strong dependence of the inverse
half-life on the value of $g_A$, since it is determined by an exponent
equal to 4, and the introduction of a quenching factor $q$ may
drastically reduce the probability of observing the \zbb~ decay.

These considerations indicate that, in order to provide reliable
calculations of the nuclear matrix elements involved in \zbb~ decay,
we need to achieve a robust knowledge of the renormalization
mechanisms of decay operators that is grounded on a microscopic
approach.
To this end, since the \zbb~ decay is ruled by the GT
spin-isospin-dependent operator, a test for the predictiveness of
nuclear structure calculations is the reproduction of observables such
as $\beta$-decay amplitudes with neutrino emission, or a GT-strength
distribution which can be obtained experimentally by way of
intermediate-energy charge-exchange reactions.

The goal of deriving effective decay operators, that account for the
degrees of freedom that have not been explicitly included in the
reduced model space, may be reached within the nuclear shell model
(SM) by resorting to the many-body perturbation theory.

Arima, Towner, and their collaborators were forerunners in the study
of the derivation of effective spin- and spin-isospin dependent
operators \cite{Arima73,Arima74,Towner83}, exploiting both sources of
the renormalization of shell-model $M1$ and GT transition operators,
namely the role of meson-exchange currents as well as the derivation
of effective operators accounting for the truncation of the Hilbert
space to the SM space \cite{Towner87,Arima88}.

In this regard, it is worth mentioning the pioneering works of Kuo and
his coworkers who, apart from the systematic development of the theory
of the effective shell-model Hamiltonian (\heff)
\cite{Kuo66,Kuo71,Kuo90}, considered for the first time the derivation
of effective shell-model \zbb-decay operators and Hamiltonians
starting from Paris \cite{Lacombe80} and Reid \cite{Reid68}
nucleon-nucleon ($NN$) potentials \cite{Wu85,Song91}.

Some of the authors of the present work started a few years ago a
systematic study of the renormalization of the GT operator accounting
for the reduced SM model space, but without considering the
corrections arising from two-body electroweak currents
\cite{Towner83,Baroni16a}. 
Our theoretical framework has been the many-body perturbation theory
\cite{Kuo81,Suzuki95,Coraggio12a,Coraggio20c}, and effective
shell-model GT operators and Hamiltonians for nuclei with mass ranging
from $A=48$ to $A=136$ have been derived starting from the
high-precision $NN$ potential CD-Bonn \cite{Machleidt01b}, whose
repulsive high-momentum components have been renormalized by way of
the so-called \vlwk~ approach \cite{Bogner02}.
The effective SM Hamiltonians and decay operators have reproduced
quantitatively the spectroscopic and decay properties, such as the
running sums of the GT strengths and nuclear matrix elements of the
\dbb~ decay (\nmed), without resorting to any quenching factor $q$,
thus indicating the reliability of our theoretical approach
\cite{Coraggio17a,Coraggio19a,Coraggio22a}.

Now, in the present paper, we report on a similar study of the
derivation of the effective SM GT decay operator \thetaeff, but
considering also the effect of the two-body electroweak currents to
ascertain the relative weight of the latter with respect to the
renormalization which accounts for the truncation of the full Hilbert
space to the reduced model space.
To this end, we start from chiral perturbation theory (ChPT), both for
the nuclear Hamiltonian \cite{Epelbaum09,Machleidt11} as well as for
the expansion of the electroweak currents which account for the
composite structure of the nucleons
\cite{Park93,Pastore09,Baroni16b}.
This leads to a consistent way to construct the effective SM operators
that are needed to construct the nuclear wave functions and calculate
then the matrix elements of GT transitions.
This approach has been extensively applied to light nuclear systems
\cite{King20,Baroni21,Gnech21,Gnech22,King23}, and recently employed
also to calculate the GT-decay strength for a few medium-mass nuclei
in terms of {\it ab initio} methods \cite{Gysbers19}.

The nuclear systems under our investigation, $^{48}$Ca, ${76}$Ge, and
${82}$Se, are candidates for the observation of \zbb~ decay.
More precisely, we present here the results of the calculations of
their GT-strength distributions and nuclear matrix elements of \dbb
decay, besides their low-energy spectroscopic properties, to validate
the quality of our calculated nuclear wave functions.
We have also performed the calculation of a large number of nuclear
matrix elements of pure GT transitions between nuclei belonging to the
$0f1p$-shell region, and compared them with the experimental ones
extracted from the data of the observed log$ft$ values.

As mentioned before, we start from a nuclear Hamiltonian based on ChPT
\cite{Epelbaum09,Machleidt11}, that consists of a high-precision
two-nucleon (2N) potential derived at next-to-next-to-next-to-leading
order (N$^3$LO) \cite{Entem02}, and a three-nucleon (3N) component at
N$^2$LO in ChPT \cite{Navratil07a}.
The one- and two-body matrix elements of the axial currents have been
derived through a chiral expansion up to N3LO, and the low-energy
constants (LECs) appearing in their expression are consistent with
those of the nuclear potential we are starting from \cite{Entem02}.

This is the first time a consistent treatment of the nuclear
Hamiltonian and of the electroweak currents has been carried out,
within the many-body perturbation theory, for nuclei that are
\zbb-decay candidates.

This paper is organized as follows.
\noindent
In Sec. \ref{outline}, first we sketch out briefly the nuclear
Hamiltonian and electroweak currents we have started from, then we
describe the perturbative approach to the derivation of the effective
SM Hamiltonian and decay operators, that we have obtained considering
$^{40}$Ca and $^{56}$Ni as doubly closed cores and the $0f1p$ and
$0f_{5/2}1p0g_{9/2}$ orbitals as model spaces, respectively.

The results of the shell-model calculations are reported in Sec. \ref{results}. 
First, we compare the calculated low-energy excitation spectra of
parent and granddaughter nuclei involved in the double-$\beta$ decays
under consideration with the experimental counterparts.
Since \heff~ that we consider for the $0f1p$ shell was extensively
investigated in a previous work \cite{Ma19}, we validate also the new
one for $0f_{5/2}1p0g_{9/2}$ model space by comparing the calculated
and experimental yrast $J^{\pi}=2^+$ excitation energies and
two-neutron separation energies ($S_{2n}$) for nickel isotopes up to
$N=48$.
Then, we report the results of the \dbb-decay matrix elements and GT
transition-strength distributions for $^{48}$Ca, $^{76}$Ge, and
$^{82}$Se, as well as of nuclear matrix elements of GT transitions for
about 40 nuclei belonging to the $0f1p$ shell.

Finally, in Sec. \ref{conclusions} we summarize the conclusions of
this study and also the outlook of our current research project.

\section{Theoretical framework}
\label{outline}
\subsection{The chiral nuclear Hamiltonian and electroweak currents}
\label{h-currents}
In the middle of the 1990s it was shown that chiral effective field
theory (ChEFT) can provide a valuable tool to deal with hadronic
interactions in a low-energy regime -- like that of nuclear systems --
with a systematic and model-independent approach
\cite{Epelbaum09,Machleidt11}. 
As is well known, one has to start identifying a clear separation of
scales \cite{vanKolck99}, and for finite nuclei we can set the pion
mass as the soft scale, $Q \sim m_\pi$, and the $\rho$ mass as the
hard scale, $\Lambda_\chi \sim m_\rho \sim 1$ GeV, which is also known
as the chiral-symmetry breaking scale.

This is the starting point of a low-energy expansion arranged in terms
of the soft scale over the hard scale, $(Q/\Lambda_\chi)^\nu$, where
$Q$ stands for an external momentum (nucleon three-momentum or pion
four-momentum) or a pion mass, and the degrees of freedom are pions
and nucleons and, eventually, their resonances ($\Delta$).

The relevant feature of ChEFT is the link with its underlying theory,
namely quantum chromodynamics; that is, the requirement to observe all
relevant symmetries of QCD, specifically the broken chiral symmetry at
low energies \cite{Weinberg79}.

In this work, we consider the high-precision $NN$ potential developed
by Entem and Machleidt, by way of a chiral perturbative expansion at
N$^3$LO \cite{Entem03}, that is characterized by a regulator function
whose cutoff parameter is  $\Lambda = 500$ MeV.
An important advantage of the EFT approach to the derivation of a
nuclear Hamiltonian is that it creates two- and many-body forces on an
equal footing \cite{Weinberg92,vanKolck94,Machleidt11}, since most
interaction vertices that appear in the three-nucleon force (3NF) and
in the four-nucleon force (4NF) also occur in the two-nucleon one
(2NF).

It is worthwhile to point out that the first nonvanishing 3NF occurs
at N$^2$LO.
At this order, there are three 3NF topologies: the two-pion exchange
(2PE), one-pion exchange (1PE), and three-nucleon-contact
interactions.
These terms are characterized by a set of low-energy constants (LECs):
the 2PE contains the parameters $c_1$, $c_3$, and $c_4$ which,
however, appear already in the 2PE component of the 2NF.

The 3NF 1PE contribution, apart from the parameters $g_A = 1.2723$,
$f_\pi=92.4$ MeV, $m_\pi=138.04$ MeV, and $\Lambda_\chi=700$ MeV,
contains a new LEC $c_D$, while another new one, $c_E$ , characterizes
the 3N contact potential.
These LECs, $c_D$ and $c_E$ , do not appear in the two-nucleon
problem, and therefore they should be fixed to reproduce the
observables of the $A=3$ system.
In present work, we have adopted the same $c_D, c_E$ values as in
Refs. \cite{Fukui18,Ma19,Coraggio20e,Coraggio21}, namely namely
$c_D=-1$ and $c_E=-0.34$.
This is a choice adopted in no-core shell model (NCSM) calculations in
Ref. \cite{Navratil07a}, where the authors first constrained the
relation of  $c_D$-$c_E$, and then investigated a set of observables
in light $p$-shell nuclei to give a second constraint.

Another innovative feature of ChPT is the possibility of constructing
electroweak currents, which account for the composite structure of the
nucleons, by way of a perturbative expansion that is consistent with
the derivation of the nuclear Hamiltonian we have just discussed
\cite{Park93,Pastore09,Kolling09,Baroni16b,Krebs17,Krebs20}.

This means that the one- and two-body matrix elements of the axial
currents ${\mathbf J_A}$ are derived through a chiral expansion up to
N$^3$LO, where the LECs appearing in their expression are consistent
with those of the nuclear potential we have considered
\cite{Entem02}.
The details about the derivation of the axial currents within chiral
effective theory can be found in Ref. \cite{Baroni16b}.

\begin{figure}[ht]
\begin{center}
\includegraphics[scale=0.60,angle=0]{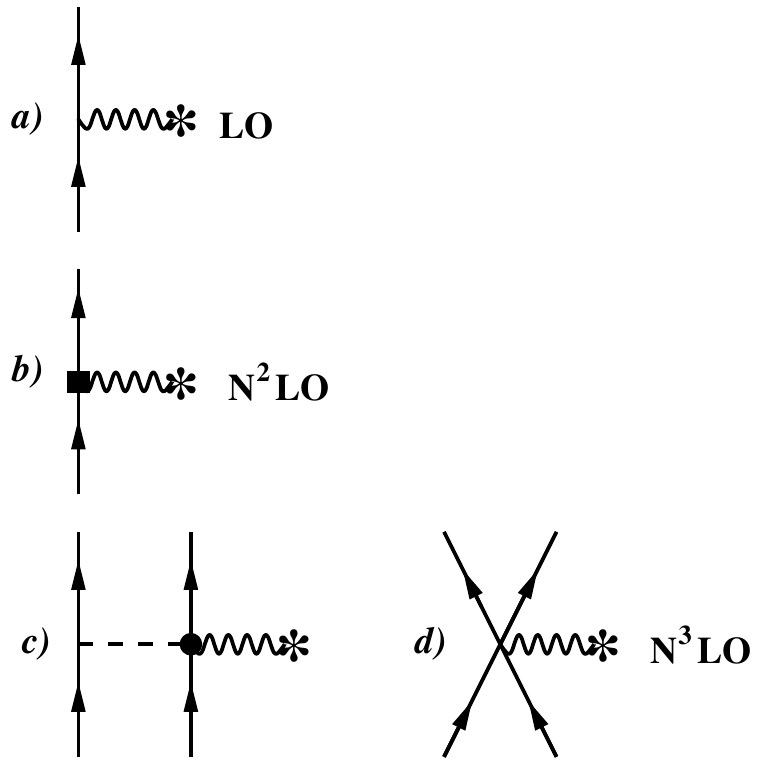}
\caption{Diagrams illustrating the contributions up to N$^3$LO to the
  axial current we have considered in the present work. The wavy lines
  represent the external weak field, the dashed lines the pion
  exchange, the square in diagram (b) represents relativistic
  corrections, while the dot in diagram (c) denotes a vertex induced by
  subleading terms in the $\pi$-nucleon chiral Lagrangian.}
\label{diagramewc}
\end{center}
\end{figure}

Here, we report the expression of ${\mathbf J_A}$ up to N$^3$LO in the
limit of vanishing momentum transfer, while its diagrammatic expansion
is illustrated in Fig. Fig. \ref{diagramewc}.

The expansion of the electroweak current ${\mathbf J_A}$ up to N$^3$LO
contains one- and two-body contributions, and therefore can be written
as

\begin{equation}
{\mathbf J_A}={\mathbf   J}_{A}(1b)+{\mathbf   J}_{A}(2b) ~~.
\end{equation}

The one-body contributions to ${\mathbf J_A}$ appear at the LO and
N$^2$LO of the ChPT expansion. 
The LO term, shown in Fig. Fig. \ref{diagramewc}$(a)$, is the standard
GT operator (with a minus sign) given by

\begin{equation}
{\mathbf   J}_{A,\pm}^{\rm LO}=-g_A\sum_i \boldsymbol{\sigma}_i
\tau_{i,\pm}~,
\end{equation}

\noindent
where $g_A=1.2723$,  and ${\bold \sigma}_i$ and $\tau_i$ are the Pauli
spin and isospin operators of the $i_{th}$ nucleon, having specified
the charge-rising $(+)$ and charge-lowering $(-)$ cases defined by

\begin{equation}
\tau_{i,\pm}=(\tau_{i,x}\pm i \tau_{i,y})/2 ~~.
\end{equation}

The N$^2$LO term [see Fig. \ref{diagramewc}$(b)$] consists of a
relativistic correction to the GT operator:

\begin{equation}
{\mathbf   J}_{A,\pm}^{\rm N^2LO}=\frac{g_A}{2m_N^2}\sum_i {\mathbf
  K}_i \times \left(\boldsymbol{\sigma}_i \times {\mathbf   K}_i
\right)\tau_{i,\pm}~,
\end{equation}

\noindent
where $m_N$ is the nucleon mass and

\begin{equation}
{\mathbf   K}_i=({\mathbf   p'}_i+{\mathbf   p}_i)/2 ~,
\end{equation}

\noindent
${\mathbf p}_i$ (${\mathbf p'}_i$) being the initial (final) momentum
of the nucleon $i$.

The first two-body diagrams appear at N3LO of the ChPT expansion,
where we have the 1PE contribution [Fig. \ref{diagramewc}$(c)$] and
the contact term (CT) [Fig. \ref{diagramewc}$(d)$], as given by

\begin{eqnarray}
\label{OPEtbc}
{\mathbf   J}_{A,\pm}^{\rm N^3LO} ({\rm 1PE};{\mathbf k})
=\sum_{i<j}  \frac{g_A}{2f^2_\pi}\biggl\{ 4c_3 \tau_{j,\pm} {\mathbf
  k} +(\boldsymbol{\tau} _i \times \boldsymbol{\tau} _j)_{\pm}
  \nonumber \\
\left. \times\left[ \left(c_4+\frac{1}{4m}\right)\boldsymbol{\sigma}
  _i\times{\mathbf k} -\frac{i}{2m}{\mathbf K}_i \right]
  \right\} \boldsymbol{\sigma} _j\cdot{\mathbf k} \frac{1}{\omega^2_k}
  + (i \rightleftharpoons j)~,
\end{eqnarray}

\begin{equation}
\label{CTtbc}
{\mathbf   J}_{A,\pm}^{\rm N^3LO} ({\rm CT};{\mathbf k})=
\sum_{i<j} z_0 (\boldsymbol{\tau} _i \times \boldsymbol{\tau}
_j)_\pm(\boldsymbol{\sigma} _i \times \boldsymbol{\sigma} _j) ~,
\end{equation}

\noindent
where $\omega_k$ is defined by the relation $\omega^2_k=k^2+m^2_\pi$
and, since the external field momentum is vanishing, we have that

\begin{eqnarray}
{\mathbf   k}_i&=&({\mathbf   p'}_i-{\mathbf   p}_i)/2=-{\mathbf   k}_j={\mathbf k}~~.
\end{eqnarray}

Moreover, we have defined

\begin{eqnarray}
 (\boldsymbol{\tau} _i \times \boldsymbol{\tau}
  _j)_{\pm}&=&(\boldsymbol{\tau} _{i} \times \boldsymbol{\tau}_{j})_x 
\pm i(\boldsymbol{\tau} _{i}	\times \boldsymbol{\tau} _{j})_y ~,
\end{eqnarray}
\noindent
and
\begin{equation}
\label{z0tbc}
z_0=\frac{g_A}{2f_\pi^2m_N}\left[ -\frac{m_N}{4g_a\Lambda_\chi}c_D
  +\frac{m_N}{3}\left(c_3+2c_4\right)+\frac{1}{6}\right]~~.
\end{equation}

The above equation shows that the 1PE term [Eq. (\ref{OPEtbc})]
contains the $c_3$ and $c_4$ LECs, which determine the two-pion
exchange contribution of the chiral 3NF at N$^2$LO, while in the CT
term [Eq. (\ref{CTtbc})] there appear, along with $c_3$ and $c_4$,
also $c_D$ [Eq. (\ref{z0tbc})] that is connected to the 3NF one-pion
exchange contribution appearing at N$^2$LO for the nuclear
Hamiltonian.

Finally, the configuration-space expression for the two-body
contribution of ${\mathbf   J}_{A}$ is obtained as
\begin{equation}
{\mathbf J}_{A,\pm}(2b)=\int \frac{d \mathbf{k}}{(2\pi)^3} 
e^{i\mathbf{k}\cdot \mathbf{r_{ij}}} C_{\Lambda}(k) {\mathbf J}_{A,\pm}(2b;\mathbf{k}) ~,
\end{equation}
\noindent
where $ \mathbf{r_{ij}}=  \mathbf{r_{i}}- \mathbf{r_{j}}$,
$C_{\Lambda}(k)= e^{-(k/\Lambda)^4}$ is the regulator function, and
$K_j$ is replaced by $-i \nabla_j$ in ${\mathbf J}_{A,\pm}
(2b;\mathbf{k})$.

We have calculated the matrix elements of these two-body terms in the
harmonic-oscillator (HO) basis, consistently with the chiral
Hamiltonian, and added them to the one-body LO and N$^2$LO operators.

It is worth pointing out that, as will be shown in
Sec. \ref{effopsec}, we have included these two-body electroweak
currents explicitly and without resorting to any approximation.
This is at variance with respect to the procedure that was followed in
Ref. \cite{Menendez11}, where the chiral two-body axial currents were
included, retaining only the normal-ordered one-body contribution by
taking as reference state a Fermi-gas approximation for the core.

The convergence with respect to the truncation of the ChPT expansion
at N$^3$LO will be discussed in Sec. \ref{results}.

\subsection{The effective shell-model Hamiltonian}
\label{heffsec}

The nuclear Hamiltonian, that includes 2NF and 3NF components, is the
foundation to build up the effective SM Hamiltonian \heff~, namely the
single-particle (SP) energies and two-body matrix elements (TBMEs) of
the residual interaction, which are the basic inputs to solve the SM
eigenvalue problem.
\heff~ must account for the degrees of freedom that are not explicitly
included in the truncated Hilbert space of the configurations (the
model space), that in our case is spanned by the proton/neutron $0f1p$
orbitals outside the doubly closed $^{40}Ca$, or by the proton/neutron
$0f_{5/2}1p0g_{9/2}$ orbitals outside the $^{56}$Ni core.

This goal may be pursued by a similarity transformation which
arranges, within the full Hilbert space of the configurations, a
decoupling of the model space $P$ where the valence nucleons are
constrained from its complement $Q=1–P$.

This problem may be tackled within the time-dependent perturbation
theory, namely by expressing \heff~ through the Kuo-Lee-Ratcliff
folded-diagram expansion in terms of the $\hat{Q}$-box vertex function
\cite{Kuo90,Hjorth95,Coraggio20c}.

The \qbox~ is defined in terms of the full nuclear Hamiltonian
$H=H_0+H_1$, where $H_0$ represents the unperturbed component,
obtained by the introduction of the harmonic-oscillator auxiliary
potential, and $H_I$ corresponds to the residual interaction, as

\begin{equation}
\hat{Q} (\epsilon) = P H_1 P + P H_1 Q \frac{1}{\epsilon-Q H Q} Q H_1
P ~, 
\label{qbox}
\end{equation}
\noindent
where $\epsilon$ is an energy parameter called ``starting energy''.

An exact calculation of the \qbox~ is practically impossible, but the
term $1/(\epsilon-QHQ)$ is amenable to be expanded as a power series,

\begin{equation}
\frac{1}{\epsilon - Q H Q} = \sum_{n=0}^{\infty} \frac{1}{\epsilon -Q
  H_0 Q} \left( \frac{Q H_1 Q}{\epsilon -Q H_0 Q} \right)^{n} ~,
\end{equation}

\noindent
leading to the perturbative expansion of the \qbox.
It is worth introducing a diagrammatic representation of the
$\hat{Q}$-box perturbative expansion, as a collection of irreducible
valence-linked Goldstone diagrams \cite{Kuo71}.

The \qbox~ is then employed to solve nonlinear matrix equations to
derive \heff~ by way of iterative techniques such as the
Kuo-Krenciglowa and Lee-Suzuki ones \cite{Suzuki80}, or graphical
noniterative methods \cite{Suzuki11}.
We have verified that the latter provide a faster and more stable
convergence to the solution of \heff, and these are the techniques we
have employed in present work.
In order to derive our \heffs, we include in our $\hat{Q}$-box
expansion one- and two-body Goldstone diagrams through third order in
the two-nucleon potential and up to first order in the three-nucleon
(NNN) one.
In Ref. \cite{Coraggio12a} a complete list of diagrams with $NN$ vertices can be found. 
The diagrams at first order in the $NNN$ potential, as well as their
analytical expressions, are reported in Refs. \cite{Fukui18,Ma19}.
It should be noted that these expressions are the coefficients of the
one-body and two-body terms arising from the normal-ordering
decomposition of the three-body component of a many-body Hamiltonian
\cite{HjorthJensen17}.
In Ref. \cite{Holt14}, Holt and coworkers showed that the
uncertainty linked to neglecting higher-order contributions from $NNN$
vertices (residual $NNN$ forces) is small.

An important issue that has to be stressed is the fact that the
nuclear systems we are going to investigate are characterized by many
valence nucleons, and this means that one should derive many-body
\heffs~ which depend on the number of valence particles. 
This implies that the \qbox~ should include at least contributions
from three-body diagrams accounting for the three-body interaction
induced by the 2NF between the valence nucleons and the configurations
outside the model space.
 
The tool we employ to diagonalize the SM Hamiltonian is the KSHELL
code \cite{KSHELL}, which cannot perform the diagonalization of a
three-body \heff.
Then, we derive a density-dependent two-body term from the three-body
contribution arising from the calculation of nine one-loop diagrams
(see the graphs in Fig. 8 of Ref. \cite{Coraggio20c}) at second order
in perturbation theory \cite{Polls83}.

We reported their explicit form in Ref. \cite{Ma19}, and it depends on
the unperturbed occupation density $\rho$ of the external valence line
that has been summed on.
This leads to the derivation of density-dependent \heffs~ which
account for the number of valence protons and neutrons, and differ
only in their TBMEs, since these one-loop diagrams are two-body
terms.

An extensive report about the perturbative behavior of the
$\hat{Q}$-expansion, starting from the chiral N$^3$LO potential of
Entem and Machleidt, was reported in Ref. \cite{Ma19}, which discussed
both its order-by-order convergence properties and the convergence
with respect to the dimension of the space of the intermediate states,
considering the systems with one and two valence neutrons, namely
$^{41}$Ca and $^{42}$Ca, respectively.
It is worth recalling here that the number of intermediate states is
expressed as a function of the maximum allowed excitation energy of
the intermediate states in terms of the oscillator quanta \nmax~
\cite{Coraggio12a}, and includes intermediate states with an
unperturbed excitation energy up to $E_{\rm max}=N_{\rm max} \hbar
\omega$.
The present limitation of our computing resources allows us to
include, for both $^{40}$Ca and ${56}$Ni cores, a maximum number of
intermediate states that does not exceed \nmax= 18.

Actually, in Ref. \cite{Ma19} we pointed out that while the energy
spacings of the theoretical SP spectra converge quite rapidly as a
function of the number of intermediate states, this convergence cannot
be reached with the larger value of the oscillator quanta we may
consider \nmax=18, when calculating their absolute values with respect
to the closed cores $^{40}$Ca, $^{56}$Ni.

Then, for the calculations of ground-state energies of nuclei in the
$0f1p$ region, we have fixed the SP energies of proton and neutron
$0_{f7/2}$ orbitals at -1.1 and -8.4 MeV, respectively.
For the \heffs~ that have been constructed with respect to the
$^{56}$Ni core, the SP energies of proton and neutron $1p_{3/2}$ have
been chosen to be -0.7 and -10.2 MeV, respectively.
Those values are consistent with experimental values of $^{41}$Sc and
$^{41}$Ca, with respect to $^{40}$Ca, and of $^{57}$Cu and $^{57}$Ni,
with respect to $^{56}$Ni \cite{Audi03}.

The Coulomb potential is explicitly taken into account, and summed to
the matrix elements of the NN potential.
The oscillator parameter $\hbar \omega$ we have employed to compute
the matrix elements of the $NN$ and $NNN$ potentials in the HO basis
is equal to 11 and 10 MeV for $^{40}$Ca and $^{56}$Ni cores,
respectively, according to the expression $\hbar \omega= 45
A^{-1/3}-25 A^{-2/3}$ \cite{Blomqvist68}.

\subsection{Effective shell-model transition operators}
\label{effopsec}
One of the goals of this work is the calculation of the matrix
elements of quadrupole-electric transition and GT-decay operators
$\Theta$ which are connected to measurable quantities such as $B(E2)$,
GT strengths, and the nuclear matrix element of the \dbb~decay \nmed.

The diagonalization of \heff~ provides the projections of the true
nuclear wave functions onto the chosen model space $P$; then we need
to renormalize any transition/decay operator $\Theta$ to account for
the neglected degrees of freedom corresponding to the $Q$ space.

The procedure that we apply to calculate effective SM operators is the
one introduced by Suzuki and Okamoto in Ref. \cite{Suzuki95}.
This approach to the derivation of effective transition/decay
operators \thetaeff~ is consistent with the one we have described in
the previous section to construct \heff, namely it is based on
perturbative expansion of a vertex function \tbox, analogously with
the derivation of \heff~ in terms of the \qbox. 
The details of such a procedure may be found in
Ref. \cite{Coraggio20c}, and in the following we only sketch out the
main building blocks.
First, we expand perturbatively the two energy-dependent vertex
functions,

\[
\hat{\Theta} (\epsilon) = P \Theta P + P \Theta Q
\frac{1}{\epsilon - Q H Q} Q H_1 P ~, \]
\[ \hat{\Theta} (\epsilon_1 ; \epsilon_2) = P H_1 Q
\frac{1}{\epsilon_1 - Q H Q} Q \Theta Q \frac{1}{\epsilon_2 - Q H Q} Q H_1 P ~,\]

\noindent
and their energy derivatives for $\epsilon=\epsilon_0$, $\epsilon_0$
being the unperturbed energy of the degenerate model space:

\[
\hat{\Theta}_m = \frac {1}{m!} \frac {d^m \hat{\Theta}
 (\epsilon)}{d \epsilon^m} \biggl|_{\epsilon=\epsilon_0} ~, \]
\[ \hat{\Theta}_{mn} =  \frac {1}{m! n!} \frac{d^m}{d \epsilon_1^m}
\frac{d^n}{d \epsilon_2^n}  \hat{\Theta} (\epsilon_1 ;\epsilon_2)
\biggl|_{\epsilon_1= \epsilon_0, \epsilon_2  = \epsilon_0} ~~.\]

Then, we can calculate a series of operators $\chi_n$:

\begin{eqnarray}
\chi_0 &=& (\hat{\Theta}_0 + h.c.)+ \hat{\Theta}_{00}~,  \label{chi0} \\
\chi_1 &=& (\hat{\Theta}_1\hat{Q} + h.c.) + (\hat{\Theta}_{01}\hat{Q}
+ h.c.) ~, \nonumber \\
\chi_2 &=& (\hat{\Theta}_1\hat{Q}_1 \hat{Q}+ h.c.) +
(\hat{\Theta}_{2}\hat{Q}\hat{Q} + h.c.) + \nonumber \\
~ & ~ & (\hat{\Theta}_{02}\hat{Q}\hat{Q} + h.c.)+  \hat{Q}
\hat{\Theta}_{11} \hat{Q}~, \label{chin} \\
&~~~& \cdots \nonumber
\end{eqnarray}

\noindent
The effective operator \thetaeff~ is then expressed in terms of an
expansion of the $\chi_n$ operators as follows:
\begin{equation}
\Theta_{\rm eff} = H_{\rm eff} \hat{Q}^{-1}  (\chi_0+ \chi_1 + \chi_2 +\cdots) ~~.
\label{effopexp}
\end{equation}

We arrest the $\chi_n$ series at $n=2$, and the $\hat{\Theta}$ vertex
functions are expanded up to third order in perturbation theory, and
in Refs. \cite{Coraggio18,Coraggio19a,Coraggio20a} we have tackled the
issue of the convergence of the $\chi_n$ series and of the
perturbative expansion of the \tbox, showing that this truncation is
substantially satisfying.
Figure 10 of Ref. \cite{Coraggio20c} and Fig. 1 of
Ref. \cite{Coraggio20a} show all the diagrams up to second order
appearing in the $\hat{\Theta}(\epsilon_0)$ expansion for a one-body
operator $\Theta^{\rm 1b}$ and two-body one $\Theta^{\rm 2b}$.

In the present work, the decay operators $\Theta $are the one-body electric-quadrupole ($E2$) transitions $q_{\scriptscriptstyle{p,n}}r^2 Y^2_m(\hat{r})$  (the charge $q_{\scriptscriptstyle {p,n}}$ being $e$
for protons and 0 for neutrons), and the axial electroweak currents
$\mathbf{J}_A$ that we employ to calculate the nuclear matrix elements
for GT decays.

We have shown in the previous section that, regarding the effective
operator for GT decays, electroweak currents have one- and two-body
components, and consequently the effective SM operator has the same
structure.
Moreover, the first-order term of the one-body component includes also
a normal-ordered contribution, calculated with respect to the
closed-core reference state, obtained from the two-body matrix
elements of the electroweak currents.

Our perturbative expansion of the \tbox~ includes contributions up to
third order in the many-body perturbation theory for the one-body
components (for both $E2$ transitions and GT decays), and up to second
order for the two-body component of the effective GT decay operator.

It is worth also reporting the explicit expression of the nuclear
matrix elements of the single-GT and \dbb~ decays in terms of the
axial electroweak currents:

\begin{eqnarray}
M_{\rm GT}^{1\nu} & = & \langle J_f || \mathbf{J}_A^{\scriptscriptstyle{\mathcal I}}
 || J_i \rangle ~, \label{meGT} \\
M_{\rm GT}^{2\nu} & = & \sum_n \frac{ \langle 0^+_f || 
 \mathbf{J}_A^{\scriptscriptstyle{\mathcal I}} || 1^+_n \rangle \langle 1^+_n ||
 \mathbf{J}_A^{\scriptscriptstyle{\mathcal I}} || 0^+_i \rangle }{E_n + E_0} ~~.
\label{doublebetameGT}
\end{eqnarray}

\noindent
where the superscript $\mathcal I$ indicates that we are employing the
matrix elements of either the bare or the effective GT decay
operators.
In the above equations, $E_n$ is the excitation energy of the $J^{\pi}
= 1^+_n$ intermediate state, and $E_0=\frac{1}{2}Q_{\beta\beta}(0^+)
+\Delta M$, where $Q_{\beta\beta}(0^+)$ and $\Delta M$ are the $Q$
value of the transition and the mass difference of the parent and
daughter nuclear states, respectively.
The index $n$ runs over all possible intermediate states induced by
the given transition operator.
It should be pointed out that we have not considered in our
calculations the Fermi component of the $\beta$-decay operator since
it plays a marginal role \cite{Haxton84,Elliott02} and in most
calculations is neglected altogether.

The calculation of \nmed~ has been carried out by way of the Lanczos
strength-function method \cite{Caurier05}, since this is the most
efficient way to include a number of $J^{\pi}=1^+$ intermediate states
that is sufficient to provide the needed accuracy.

Then, the calculated value of \nmed~ can be compared with the
experimental counterpart, which is extracted from the observed half
life $T^{2\nu}_{1/2}$:

\begin{equation}
\left[ T^{2\nu}_{1/2} \right]^{-1} = G^{2\nu} \left| M_{\rm GT}^{2\nu}
\right|^2 ~,
\label{2nihalflife}
\end{equation}
\noindent
$G^{2\nu}$ being the \dbb-decay phase-space (or kinematic) factor
\cite{Kotila12,Kotila13}.

\section{Results}
\label{results}
In this section we present the results of our SM calculations. 
First, to validate the quality of the nuclear wave functions we employ
to describe GT decays, we compare the theoretical low-energy
spectroscopic properties of a few nuclei, in the mass regions which
are of relevancy for the \heffs~ we have derived, with the
experimental ones.
In Refs. \cite{Ma19,Coraggio20e,Coraggio21} we have already shown the
ability of the $0f1p$-shell \heffs~ that we have derived, starting
from the chiral nuclear Hamiltonian we presented in
Sec. \ref{h-currents}, to reproduce accurately the low-energy spectra
and monopole properties of calcium, titanium, chromium, iron, and
nickel isotopes.

Therefore, in Sec. \ref{nickel} we limit our discussion only to the
\heffs we have constructed for the $0f_{5/2}1p0g_{9/2}$ model space,
focusing on their monopole properties.
It is worth emphasizing that, since we derive \heff considering the
contribution of the induced three-nucleon potential too, as reported
in Sec. \ref{heffsec}, we use a different \heff for each nuclear
system.

In the Supplemental Material \cite{supplemental2023} the effective SM
Hamiltonian for $A= 58$ systems, namely for one- and two-valence
nucleon systems, can be found.

\begin{figure}[ht]
\begin{center}
\includegraphics[scale=0.22,angle=0]{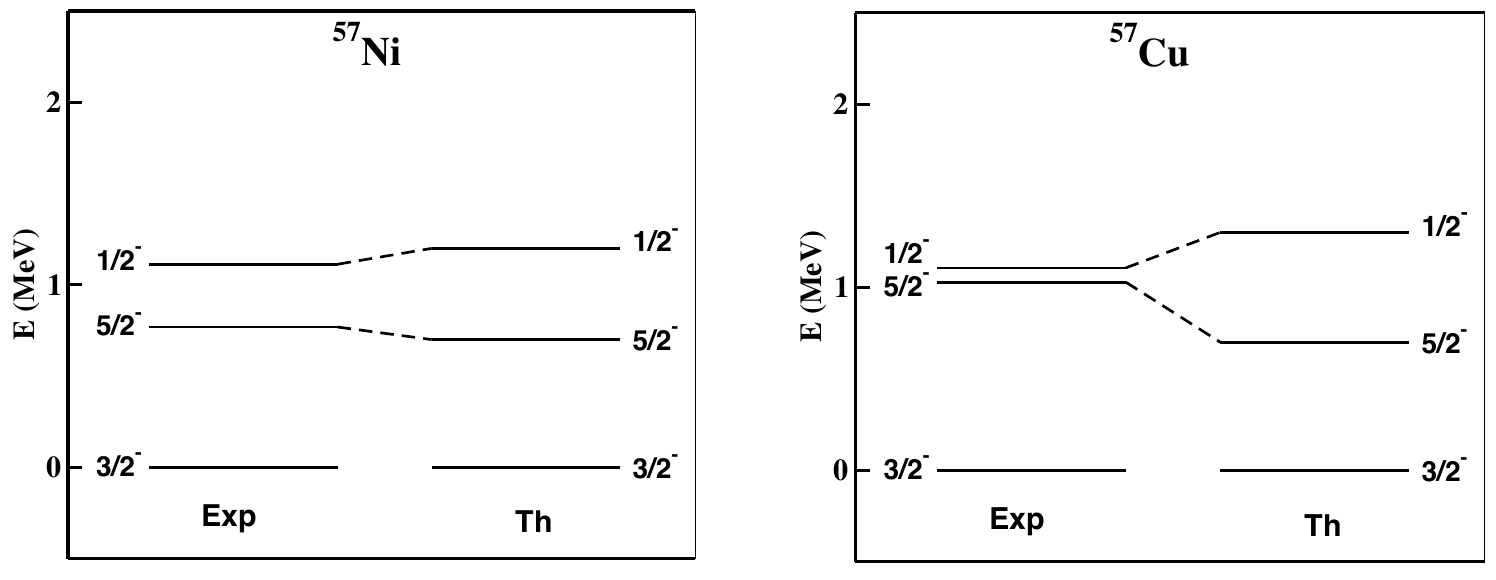}
\caption{Comparison between experimental and calculated SP spectra of
  $^{57}$Ni and $^{57}$Cu.} 
\label{57Ni57Cu}
\end{center}
\end{figure}

In Sec. \ref{spectra} we present the results of the calculation of the
low-energy spectroscopic properties of the parent and granddaughter
nuclei that are involved in the double-$\beta$ decay processes we
consider in this paper, namely $^{48}$Ca, $^{48}$Ti, $^{76}$Ge,
$^{76}$Se, $^{82}$Se, and $^{82}$Kr, and compare them with available
data.

Finally, in Sec. \ref{GT} we compare with experiment the calculated
properties related to the GT decay, such as the nuclear matrix
elements of about 40 nuclear systems in the $0f1p$-shell region, the
GT$^-$ strength distributions, and the \nmeds~ of the above mentioned
nuclei.

\subsection{Calculation of the spectroscopic properties}
\label{spectroscopy}

\subsubsection{Monopole properties of \heff~in the $0f_{5/2}1p0g_{9/2}$
  model space}
\label{nickel}
As previously mentioned, the effective SM Hamiltonian for $A=58$
nuclei can be found in the Supplemental Material of this paper, and
the values of our calculated SP energies are $\epsilon_{1p3/2}=0.0$
MeV, $\epsilon_{0f5/2}=0.7$ MeV, $\epsilon_{1p1/2}=1.3$ MeV, and
$\epsilon_{0g/2}=6.2$ MeV for the proton orbitals, and
$\epsilon_{1p3/2}=0.0$ MeV, $\epsilon_{0f5/2}=0.7$ MeV,
$\epsilon_{1p1/2}=1.2$ MeV, and $\epsilon_{0g/2}=6.1$ MeV for the
neutron orbitals.

Since the SP energies are the eigenvalues of the \heffs~ for the
single-valence-nucleon systems, from the SM perspective these numbers
correspond to the excitation energies of the SP states in $^{57}$Cu
and $^{57}$Ni.
In Fig. \ref{57Ni57Cu} we compare the theoretical and experimental SP
spectra of these two nuclei \cite{ensdf}; note that there is no firm
assignation of a $J^{\pi}=9/2^+$ SP state for either of them.

From the inspection of Fig. \ref{57Ni57Cu}, we observe that the
calculated SP energies reproduce quite well the observed
natural-parity SP spacings.
As a test case of the isotopic chains belonging to the
$0f_{5/2}1p0g_{9/2}$ model space, we examine the nickel isotopes,
whose study is pivotal to investigate the shell-closure properties of
our calculated \heffs.

\begin{figure}[ht]
\begin{center}
\includegraphics[scale=0.31,angle=0]{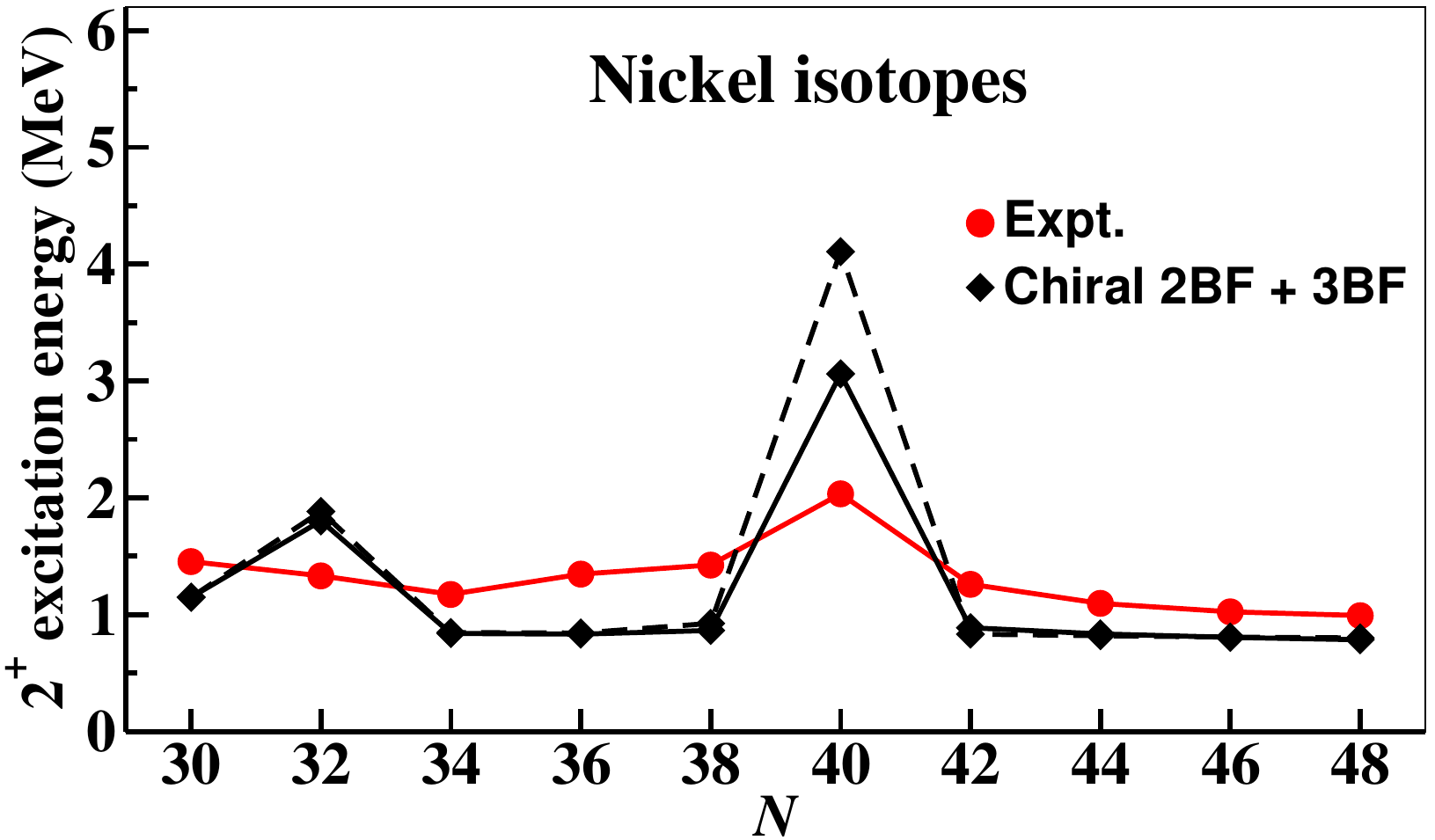}
\caption{Experimental and calculated excitation energies of the yrast
  $J^{\pi}=2^+$  states for nickel isotopes from $N = 30$ to 48. The
  black dashed line refers to results obtained with the \heff~ for the
  $A=58$ system (see text for details).}
\label{J2pNi}
\end{center}
\end{figure}

In Fig. \ref{J2pNi} we show the behavior of the calculated and
experimental $J^{\pi}=2_1^+$ excitation energies of nickel isotopes up
to $N=48$ \cite{ensdf}.
The black dashed line indicates the results obtained with the $A=58$
effective Hamiltonian (see the Supplemental Material
\cite{supplemental2023}), and the black solid line refers to the
results obtained with density-dependent \heffs~ accounting for the
induced three-body potential (see Sec. \ref{heffsec}).

As can be seen, the theoretical results follow quite well the observed
behavior of the yrast $J^{\pi}=2_1^+$ states all along the isotopic
chain, except for the energy bump at $N=32$ which is a fingerprint of
the subshell closure of the neutron $1p_{3/2}$ orbital, thus revealing
that the $Z=28$ cross-shell excitations still play an important role
in describing the low-energy spectroscopy of light nickel isotopes
\cite{Coraggio14b}.

It should be noted that our SM calculations reproduce the shell
closure at $N=40$ ($^{68}$Ni), the result being remarkably better by
employing the density-dependent \heffs.
This confirms the ability of the monopole component of our \heffs~ to
provide the observed shell evolution, a feature that may be ascribed
to the $NNN$ component of the chiral nuclear Hamiltonian, as we showed
in our previous study of $0f1p$-shell nuclei \cite{Ma19}.

\begin{figure}[H]
\begin{center}
\includegraphics[scale=0.31,angle=0]{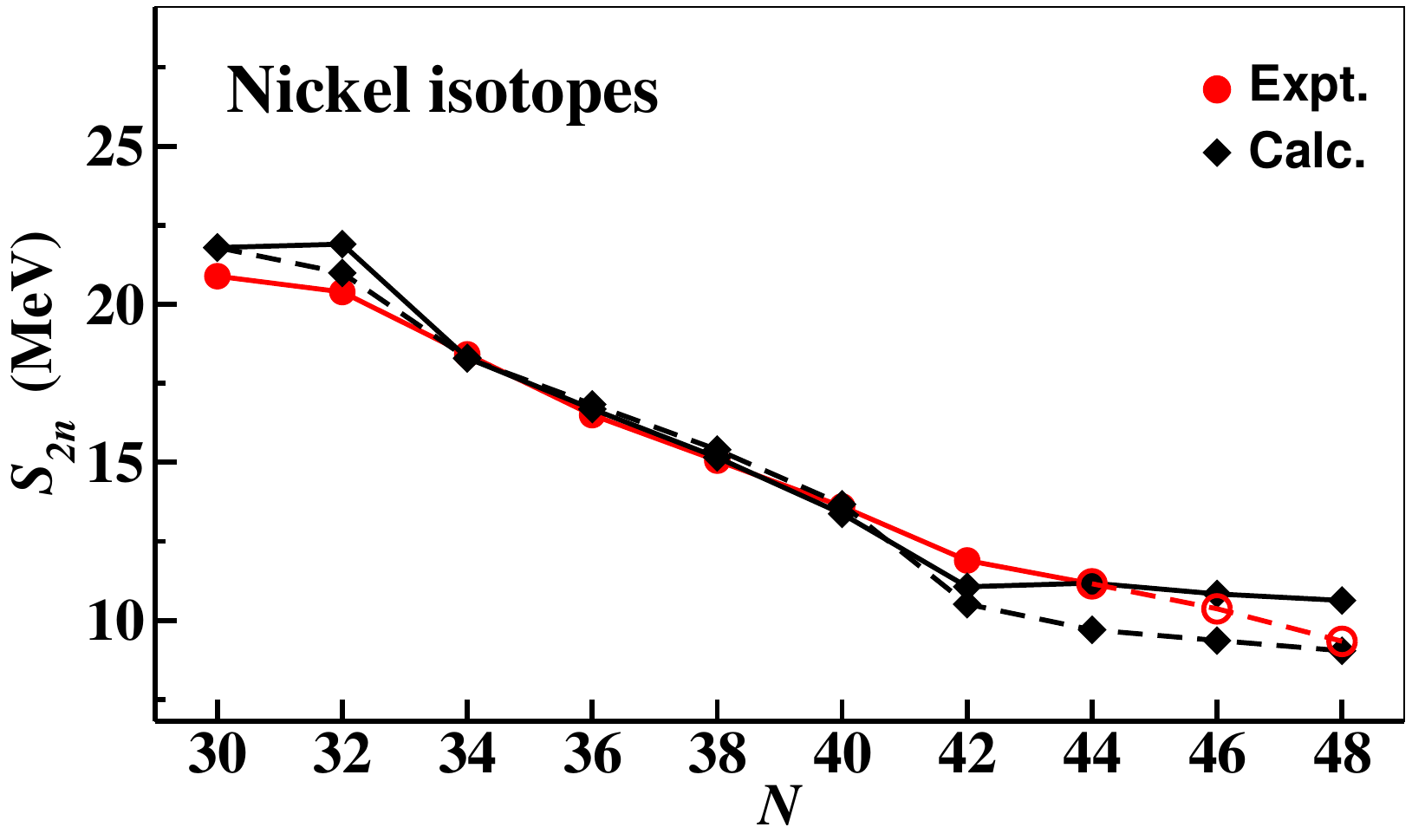}
\caption{Experimental and calculated two-neutron separation energies
  for nickel isotopes from $N = 30$ to 48. The black dashed line
  refers to results obtained with the \heff~ for the $A=58$ system (see
  text for details). Data are taken from \cite{Audi03}; open circles
  correspond to estimated values.}
\label{S2nNi}
\end{center}
\end{figure}

Similar positive conclusions may be drawn from the inspection of the
behavior of the nickel two-neutron separation energies ($S_{2n}$) as a
function of the neutron number, which we have reported in
Fig. \ref{S2nNi}.
We note that the discrepancy between the results obtained with $A=58$
\heff~ (dashed black line) and the density-dependent \heffs~ (solid
black line) starts to enlarge at $N=44$, but from that point on there
are only extrapolated values and no experimental counterpart.

\begin{figure}[H]
\begin{center}
\includegraphics[scale=0.30,angle=0]{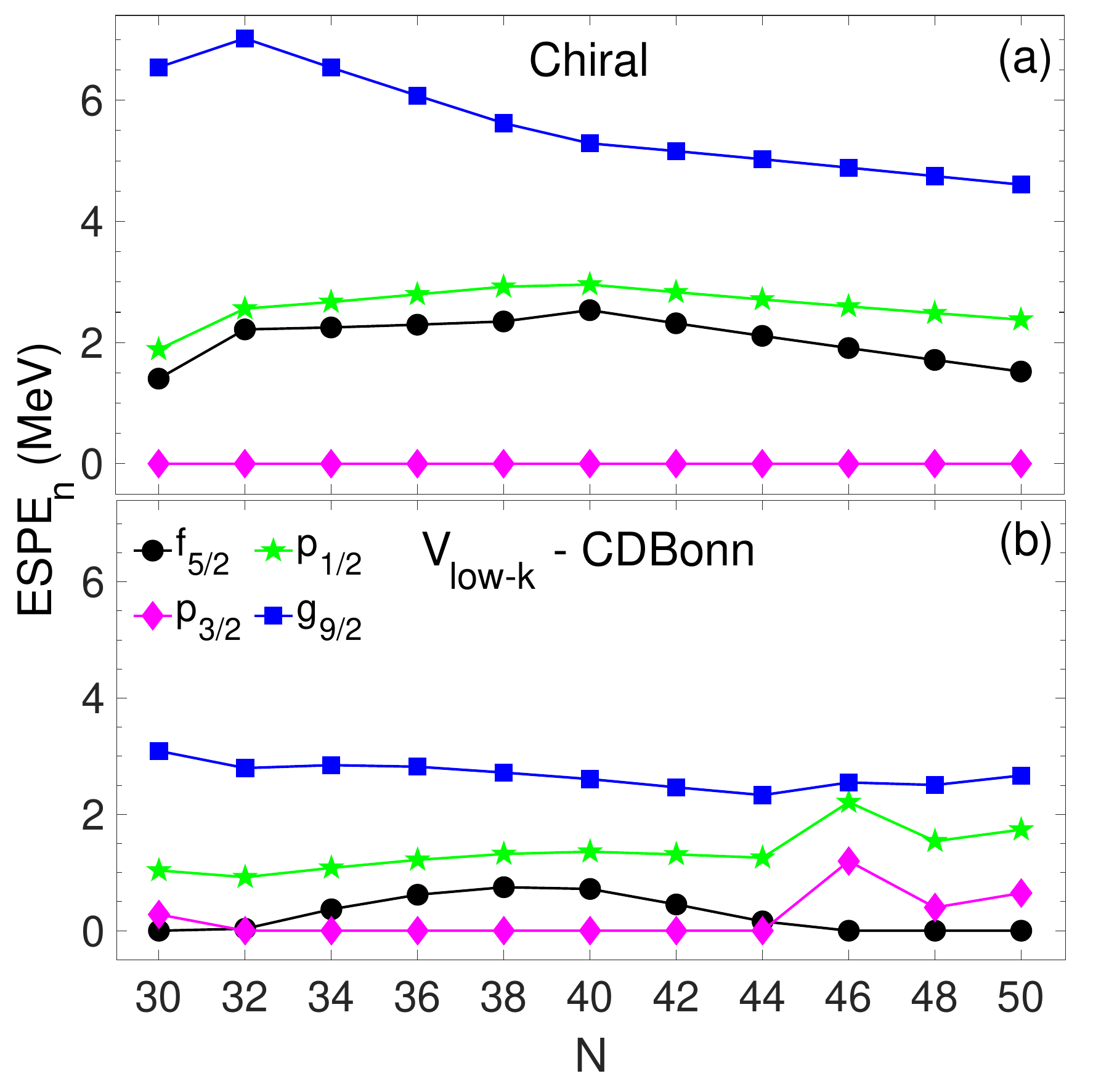}
\caption{Neutron ESPEs from \heff~ TBMEs for nickel isotopes as a
  function of the neutron number, calculated from the CD-Bonn and from
  chiral two- and three-body potential (see the text for details).}
\label{espe_Nickel}
\end{center}
\end{figure}

We have pointed out in our previous study of nuclear systems belonging
to the $0f1p$-shell region that a major role in driving the shell
evolution, starting from chiral ChPT nuclear Hamiltonians, is played
by 3NF that contribute significantly to the monopole component of
\heff~ \cite{Ma19}.
The direct link between the properties of the monopole component and
the shell evolution is provided by the calculation and the study of
the effective single-particle energy (ESPE) that is defined in terms
of the bare SP energy $\epsilon_j$ and the monopole part of the TBMEs
\cite{Utsuno99,Otsuka22}:

\begin{equation}
  {\rm ESPE}(j) = \epsilon_j + \sum_{j'} V^{mon}_{j j'} n_{j'}~,
\label{eqespe}
\end{equation}

\noindent
where the sum runs over the model-space levels $j'$, $n_{j}$ being the
occupation number of particles in the level $j$ obtained from the
diagonalization of the shell-model Hamiltonian, and the
angular-momentum-averaged monopole component of the SM Hamiltonian is
defined through the TBMEs of the SM residual interaction $V_{\rm eff}$
as follows:

\[
V^{mon}_{ij}=\frac{\sum_J  (2J +1) \langle i,j |V_{\rm eff} | i,j \rangle_J}{
                                                    \sum_J (2J +1)} ~~.
\]

To illustrate the connection between the evolution of the ESPEs as a
function of the valence nucleons and the closure properties of \heffs,
in Fig. \ref{espe_Nickel} we compare the neutron ESPEs of nickel
isotopes obtained with two different \heffs, both of them defined in
the $0f_{5/2}1p0g_{9/2}$ model space: the \heff we have derived for
the present work from chiral 2NF and 3NF, and that which was
constructed and employed in Ref. \cite{Coraggio19a} starting from the
CD-Bonn high-precision NN potential \cite{Machleidt01b} renormalized
by way of the \vlwk~ procedure \cite{Bogner02}.
Black dots, magenta diamonds, green stars, and blue squares indicate
the $0f_{5/2}$, $1p_{3/2}$, $1p_{1/2}$, and $0g_{9/2}$ ESPEs,
respectively.

\begin{figure}[ht]
\begin{center}
\includegraphics[scale=0.31,angle=0]{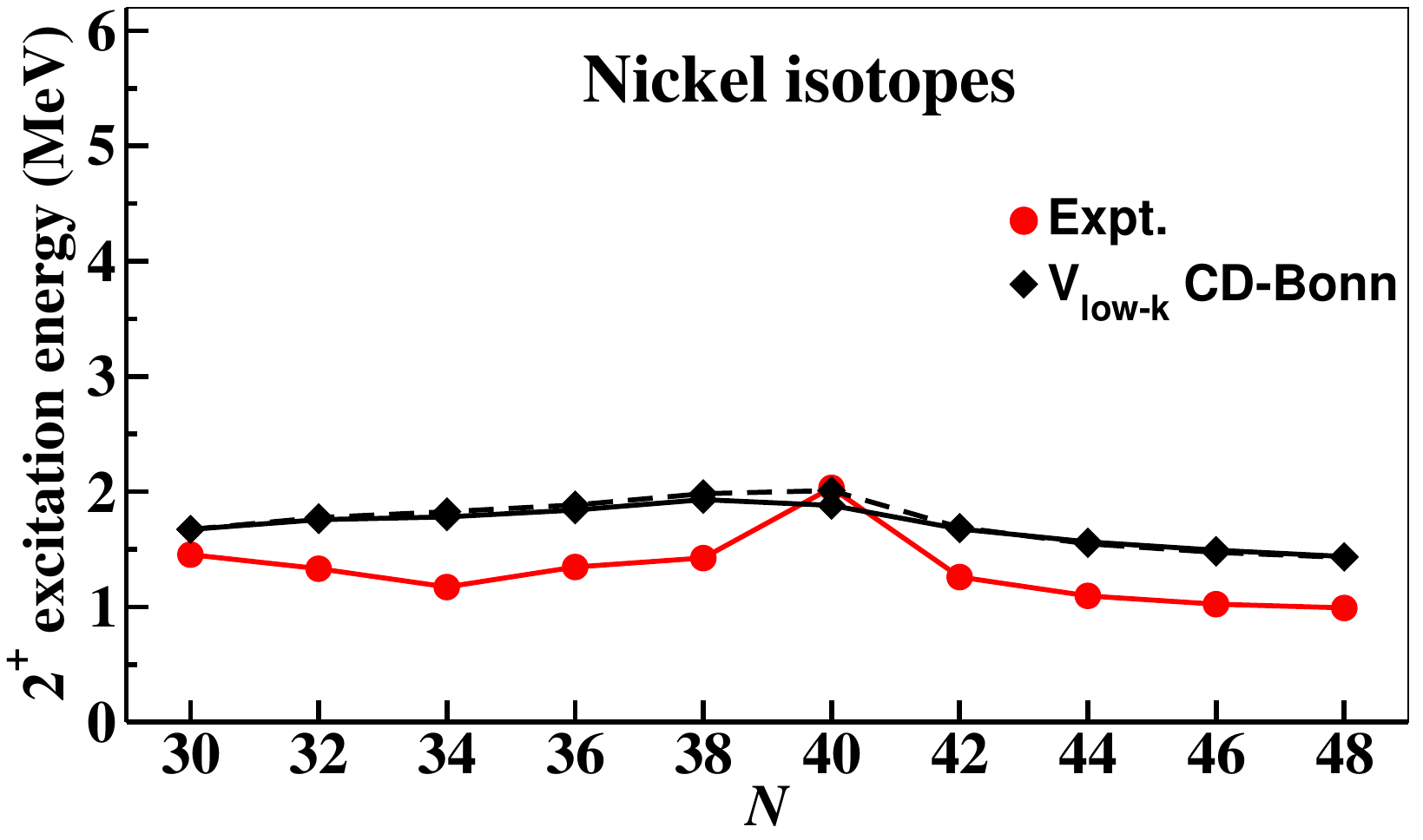}
\caption{Same as in Fig. \ref{J2pNi}, but the theoretical results are
  here obtained with \heff~reported in Ref. \cite{Coraggio19a} (see
  text for details).}
\label{J2pNi-vlwk}
\end{center}
\end{figure}

The inspection of Fig. \ref{espe_Nickel} shows that the \heff from the
ChPT Hamiltonian generates ESPEs that are characterized by an almost
constant energy splitting along the isotopic chain, the $0g_{9/2}$
ESPE being always well separated from the other ones.
This behavior of the ESPEs is reflected in the neutron closure at $N =
40$, as emerges from the results in Fig. \ref{J2pNi}.

The situation is different when considering the ESPEs obtained from
the \heff~ that was employed in Ref. \cite{Coraggio19a}, where the
energy splittings are strongly reduced, exposing then the shell
evolution to the correlations induced by higher-multipole components
of the residual two-body potential.
This feature is reflected by a collective flat behavior of the
calculated excitation energies of the yrast $J^{\pi}=2^+$ states, as
can be seen in Fig. \ref{J2pNi-vlwk}, reaching the climax with the
disappearance $N=40$ closure.

\subsubsection{Low-energy spectra of $^{48}$Ca, $^{48}$Ti,
$^{76}$Ge, $^{76}$Se, $^{82}$Se, and $^{82}$Kr}
\label{spectra}
Before starting the analysis of the calculated quantities related to
the GT decay, we deem it is worth comparing our calculated low-energy
spectra of $^{48}$Ca, $^{48}$Ti, $^{76}$Ge, $^{76}$Se, $^{82}$Se, and
$^{82}$Kr, as well as their electromagnetic-transition properties,
with the available experimental counterparts.

All calculations have been performed employing theoretical SP
energies, TBMEs, and effective transition operators, following the
procedure described in Secs. \ref{heffsec},\ref{effopsec}.
The TBMEs include the contribution of induced three-body forces, so
they depend on the different nuclear system under consideration (see
the content in Sec. \ref{heffsec}).

\begin{center}
\begin{figure}[ht]
\includegraphics[scale=0.32,angle=0]{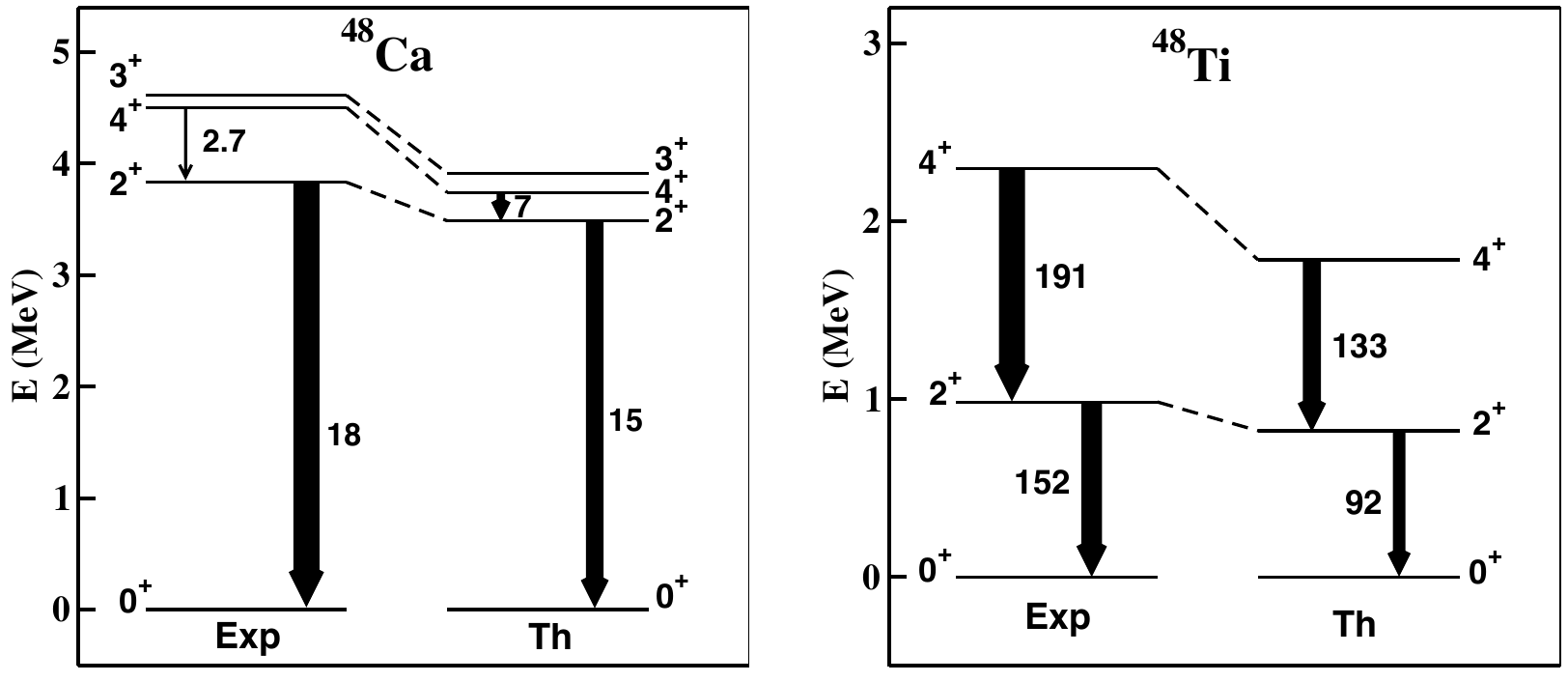}
\caption{Experimental and calculated spectra of $^{48}$Ca and
  $^{48}$Ti. $B(E2)$ strengths (in $e^2{\rm fm}^4$) are also reported
  (see text for details).}
\label{48Ca48Ti}
\end{figure}
\end{center}

The shell-model calculation for $^{48}$Ca and $^{48}$Ti are performed
within the full $fp$ shell, namely the proton and neutron $0f_{7/2}$,
$0f_{5/2}$, $1p_{3/2}$, and $1p_{1/2}$ orbitals.
In Fig. \ref{48Ca48Ti}, we show the experimental \cite{ensdf} and
calculated low-energy spectra of $^{48}$Ca and $^{48}$Ti.
Next to the arrows, whose widths are proportional to the $B(E2)$
strengths, we report the numerical values in \ref{48Ca48Ti}
\cite{ensdf}.

As can be seen, we reproduce very nicely the observed shell closure of
the neutron $0f_{7/2}$ orbital in $^{48}$Ca, the agreement between the
experimental and calculated spectra may be considered quantitative,
and also the observed $B(E2)$s are satisfactorily reproduced by the
theory.

\begin{center}
\begin{figure}[H]
\includegraphics[scale=0.32,angle=0]{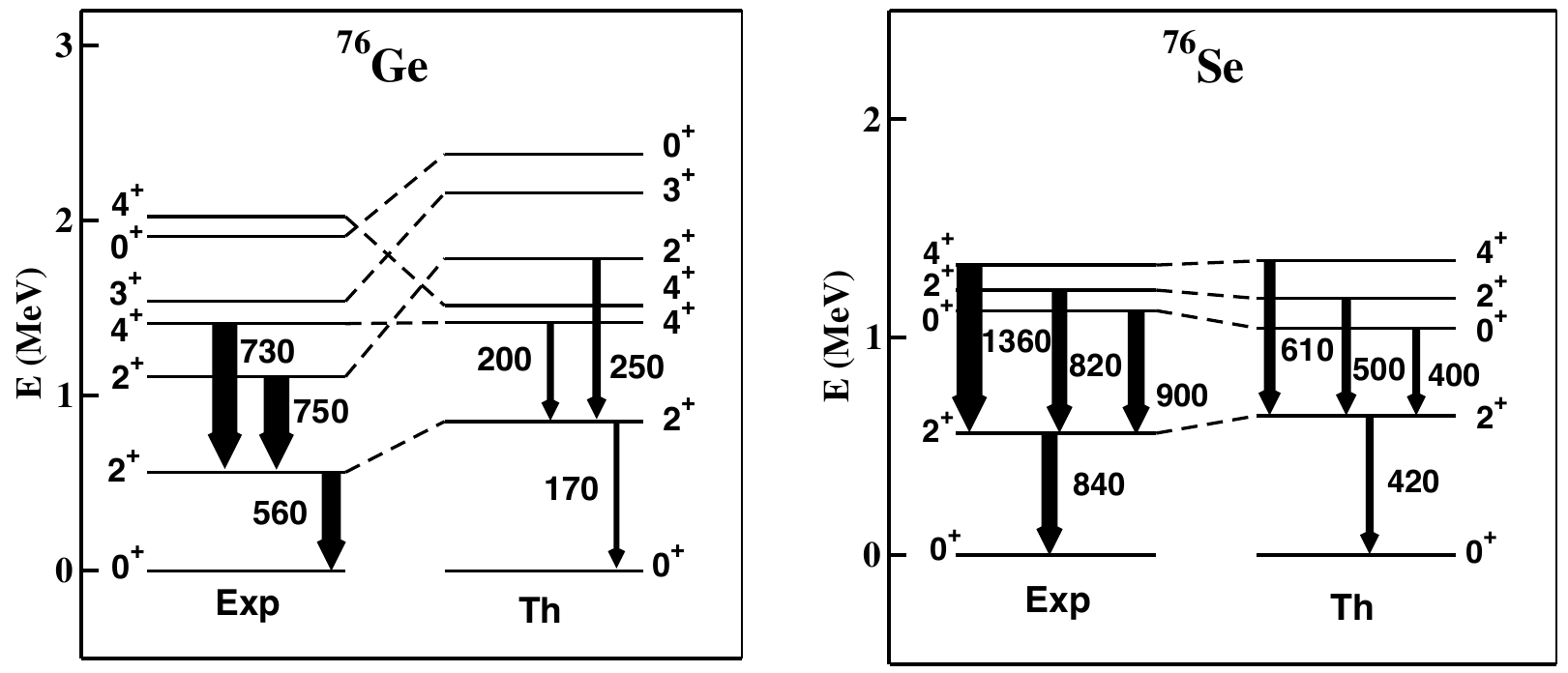}
\caption{ Same as in Fig. \ref{48Ca48Ti}, but for $^{76}$Ge and
  $^{76}$Se (see text for details).}
\label{76Ge76Se}
\end{figure}
\end{center}

As already pointed out in the previous sections, the shell-model
calculations for $^{76}$Ge, $^{76}$Se, $^{82}$Se, and $^{82}$Kr have
been performed within the model space spanned by the four proton and
neutron orbitals $0f_{5/2}$, $1p_{3/2}$, $1p_{1/2}$ and $0g_{9/2}$,
considering $^{56}$Ni as a closed core.
The experimental \cite{ensdf} and calculated low-energy spectra of
$^{76}$Ge and $^{76}$Se are reported in Fig. \ref{76Ge76Se}, together
with the experimental \cite{ensdf} and calculated $B(E2)$ strengths
(in $e^2{\rm fm}^4$).

We notice that while the agreement between the experimental and
calculated spectra and $B(E2)$’s for $^{76}$Se is quite satisfactory,
the same conclusion does not apply to $^{76}$Ge, whose observed
collectivity is poorly described. 
These results are at variance with respect to those we found in
Ref. \cite{Coraggio19a}, where the \heff~ was derived from a $NN$
\vlwk~ potential obtained from the CD-Bonn potential
\cite{Machleidt01b}, and the reproduction of the $^{76}$Ge spectrum
and $B(E2)$’s was far better than the one in Fig. \ref{76Ge76Se}.

There is experimental evidence that low-energy states of $^{76}$Ge
reveal a rigid triaxial deformation \cite{Toh13}, and this enhanced
collectivity, that characterizes heavy-mass germanium isotopes, may be
reproduced within shell model by employing a model space larger than
the $0f_{5/2}1p 0g_{9/2}$0 one \cite{Rhodes22}.

The different collective behavior of the calculated $^{76}$Ge
low-lying energy spectrum, as obtained with present the chiral \heff~
and the one in Ref. \cite{Coraggio19a}, traces back to their different
monopole components.
In Table \ref{ESPE76Ge} we report the proton and neutron ESPEs for
$^{76}$Ge, calculated with both \heffs, and it is evident that the SM
Hamiltonian employed in Ref. \cite{Coraggio19a} provides an ESPE
spectrum more compressed than the one obtained with the chiral \heff.

\begin{table}[ht]
\caption{Proton and neutron ESPEs (in MeV) for $^{76}$Ge calculated
  with the \heffs~derived from the chiral 2NF and 3NF (this work), and
  from the \vlwk~potential obtained from the CD-Bonn potential
\cite{Coraggio19a}.}
\begin{ruledtabular}
\begin{tabular}{ccccccccc}
\label{ESPE76Ge}
 orbital & ~ & ~ & proton & ~ & ~ & ~ & neutron & ~ \\
\colrule
 ~ & ~ & Chiral & ~ & \vlwk ~ & ~ & Chiral & ~ & \vlwk \\
\colrule
$0f_{5/2}$ & ~ & 2.9 & ~ & 0.3 & ~ & 3.6 & ~ & 0.5 \\
$1p_{3/2}$ & ~ & 0.0 & ~ & 0.0 & ~ & 0.0 & ~ & 0.0 \\
$1p_{1/2}$ & ~ & 3.1 & ~ & 0.9 & ~ & 3.1 & ~ & 1.3 \\
$09_{9/2}$ & ~ & 7.0 & ~ & 3.3 & ~ & 6.3 & ~ & 2.5 \\
\end{tabular}
\end{ruledtabular}
\end{table}

The collectivity induced by the smaller energy spacings of the ESPEs,
that are calculated starting from the \vlwk~ renormalization of the
CD-Bonn potential, has a drawback: in a few cases this feature fails
to reproduce shell closures, such as in 48Ca (see Fig. 5 in
Ref. \cite{Coraggio19a}) or in $^{68}$Ni, as we have shown in
Fig. \ref{J2pNi-vlwk}.

\begin{center}
\begin{figure}[hb]
\includegraphics[scale=0.32,angle=0]{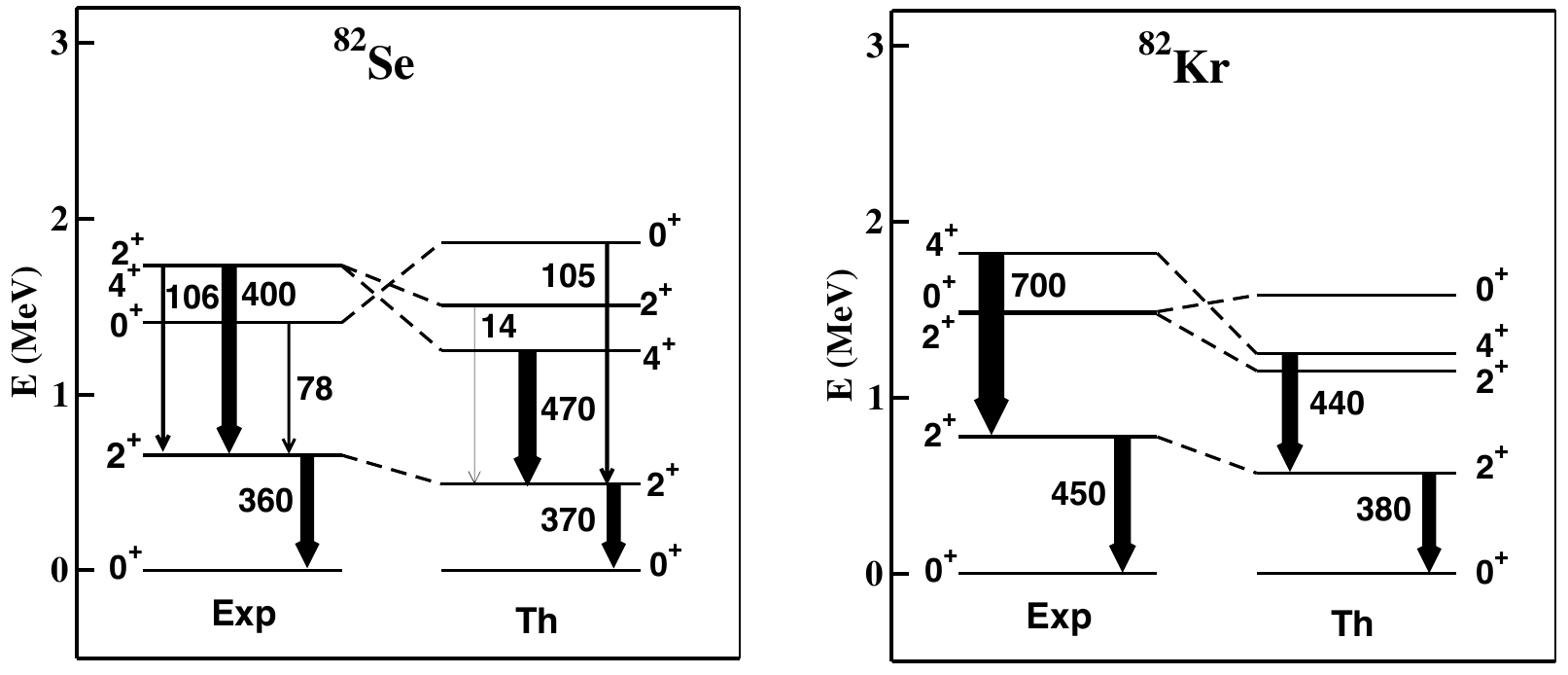}
\caption{ Same as in Fig. \ref{48Ca48Ti}, but for $^{82}$Se and
  $^{82}$Kr.}
\label{82Se82Kr}
\end{figure}
\end{center}

Also, for the shell model calculation of $^{82}$Se and $^{82}$Kr, we
have considered the $0f_{5/2}1p0g_{9/2}$ model space, and in
Fig. \ref{82Se82Kr} we report the experimental \cite{ensdf}  and
theoretical low-energy spectra and $B(E2)$’s.

There is little to comment; the agreement between theory and
experiment, especially as regards the electromagnetic properties, can
be considered very satisfactory, with the exception of the inversion
in the calculated spectra of the $0^+_2$ states with respect to the
$2^+_2$ and $4^+_1$ ones.

\subsection{Nuclear matrix elements of the GT decay}
\label{GT}
In this section, we present the results of our calculations of nuclear
matrix elements and GT$^-$ strength distributions for $^{48}$Ca,
$^{76}$Ge, and $^{82}$Se, and compare them with the available data.

\begin{center}
\begin{figure}[ht]
\includegraphics[scale=0.26,angle=0]{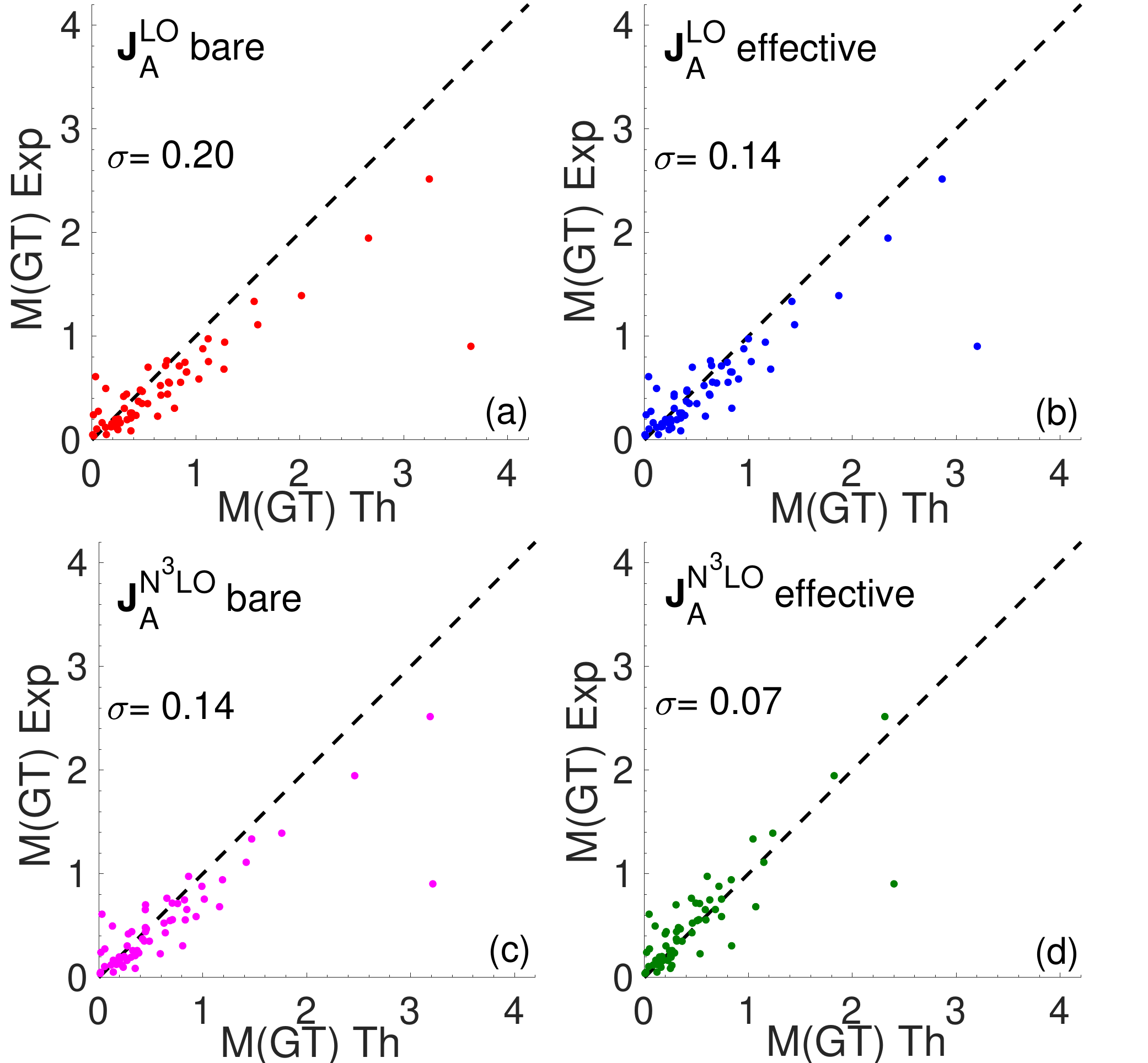}
\caption{Correlation plots between experimental (y-axis) and calculated
  (x-axis) values of the GT nuclear matrix elements of a few decay
  processes in $0f1p$ shell region. The experimental value are taken
  from Ref. \cite{ensdf}.}
\label{corrplot}
\end{figure}
\end{center}

In order to assess the distinct role that is played by the
meson-exchange currents -- that are taken into account by ChPT
expansion of the axial current -- and by the configurations outside
the model space in the renormalization of the shell-model GT-decay
operator, we label our calculations as follows:

\begin{itemize}
\item[(a)] calculations performed by employing the bare ${\mathbf J_A}$
  at LO in ChPT, namely the usual spin-isospin dependent GT operator
  $g_A \boldsymbol{\sigma} \cdot \boldsymbol{\tau}$;
\item[(b)] calculations performed by employing the effective ${\mathbf J_A}$
  at LO in ChPT, that accounts for the contributions of configurations
  outside the model space (see Sec. \ref{effopsec});
\item[(c)] calculations performed by employing the bare ${\mathbf J_A}$
  at N$^3$LO in ChPT, that includes also the contributions of the
  relativistic corrections to the GT operator and the two-body contact
  and pion-exchange contributions;
\item[(d)] calculations performed by employing the effective ${\mathbf J_A}$
  at N$^3$LO in ChPT, a SM operator that owns both one- and two-body
  components.
\end{itemize}

First, we have considered 60 experimental GT decays of 43 nuclei
belonging to the region of the $0f1p$ shell \cite{ensdf}, involving
only yrast states, and compared the matrix elements extracted from the
data with our calculated values.
We report them in a correlation plot in Fig. \ref{corrplot}, where the
theoretical nuclear matrix elements are obtained employing SM decay
operators (a)–(d), together with the root-mean-square deviation
$\sigma$:
\[
\sigma = \sqrt{\frac{\sum_{i=1}^{n} (x_i-\hat{x}_i)^2}{n}}~,
\]
\noindent
$x_i$ being the experimental GT nuclear matrix element, $\hat{x}_i$
the corresponding calculated value, and $n=60$ the total number of
data we have considered (see the table in the Supplemental Material
\cite{supplemental2023}).
This analysis is analogous to the one reported in Fig. 1 of
Ref. \cite{Martinez-Pinedo96}, where the authors evidenced the need to
introduce a quenching factor $q \approx 0.74$ of the axial coupling
constant $g_A$ to reproduce at best GT data of nuclei in the $0f1p$
region with SM eigenfunctions obtained by diagonalizing the KB3 SM
Hamiltonian \cite{Poves81}.

\begin{center}
\begin{figure}[ht]
\includegraphics[scale=0.36,angle=0]{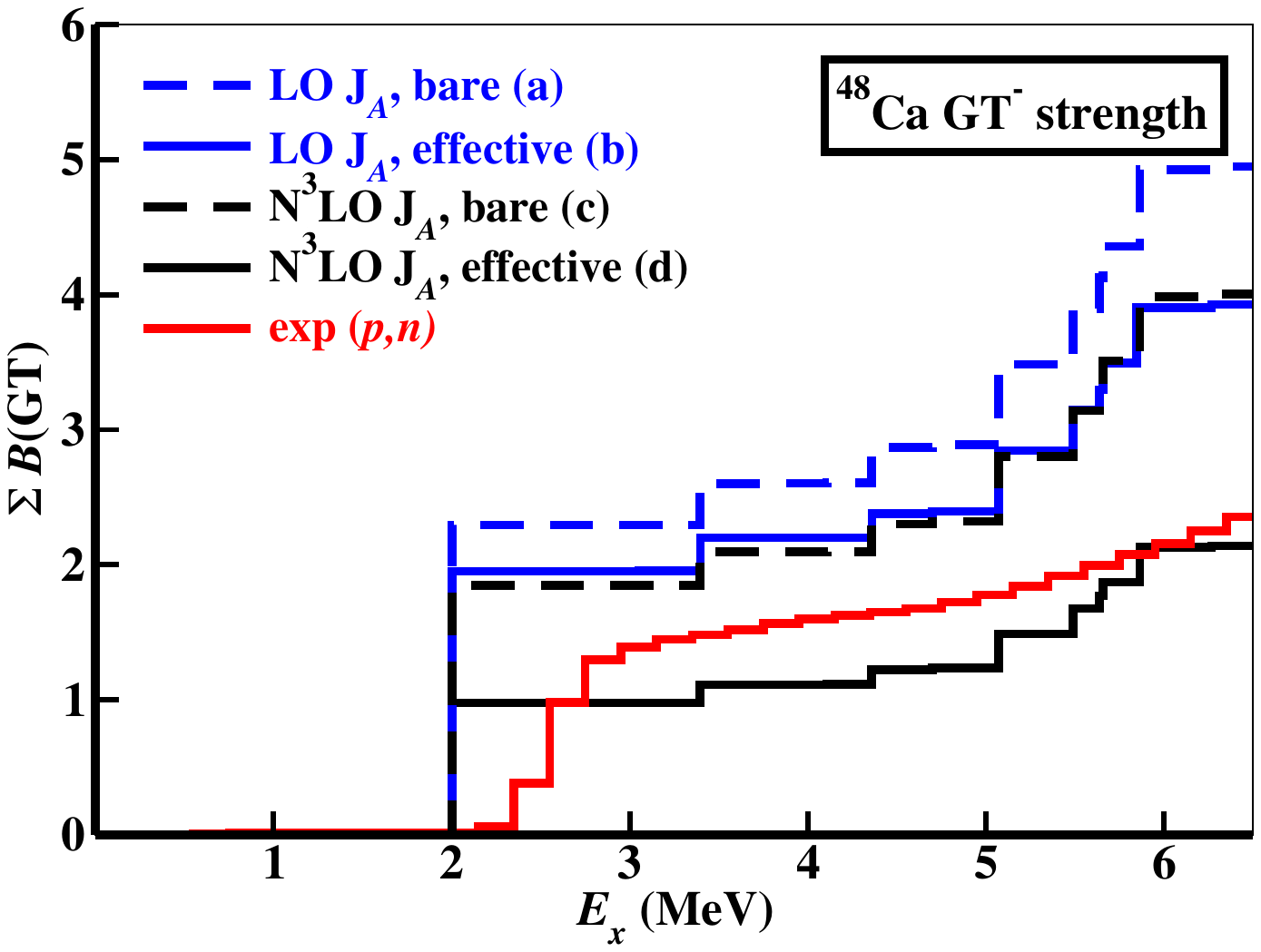}
\caption{Running sums of the $^{48}$Ca $B({\rm GT})$ strengths as a
  function of the excitation energy $E_x$ up to 6.5 MeV (see text for
  details).}
\label{48CaGT-}
\end{figure}
\end{center}

From the inspection of Fig. \ref{corrplot}, the best overall
reproduction of the data is obtained by carrying out calculations with
the effective operator (d), and the values of $\sigma$ obtained with
the operators (b) and (c) show that for the $0f1p$ shell the
improvements we obtain with respect to the results with the bare
operator (a) can be equally ascribed to the renormalization of the GT
operator by way of the ChPT expansion of ${\mathbf J_A}$, as well as
to the derivation of SM effective operators which account for the
configurations outside the model space (see Sec. \ref{effopsec}).

Another quantity that is indirectly related to the GT decay operator,
and that is worth studying, is the GT strength distribution:

\begin{equation}
B({\rm GT})= \frac{ \left| \langle \Phi_f || {\mathbf J_A}/g_A || \Phi_i \rangle \right|^2} 
{2J_i+1}~,
\label{GTstrength}
\end{equation}

\noindent
where indices $i,f$ refer to the parent and daughter nuclei, respectively.

They are obtained from charge-exchange reactions, and can be extracted
from the GT component of the cross section at zero degrees, following
the standard approach in the distorted-wave Born approximation (DWBA):

\[
\frac{d\sigma^{GT}(0^\circ)}{d\Omega} = \left (\frac{\mu}{\pi \hbar^2} \right
)^2 \frac{k_f}{k_i} N^{\sigma \tau}_{D}| J_{\sigma \tau} |^2 B({\rm
  GT})~,
\]

where $N^{\sigma \tau}_{D}$ is the distortion factor, $| J_{\sigma
  \tau} |$ is the volume integral of the effective $NN$ interaction,
$k_i$ and $k_f$ are the initial and final momenta, respectively, and
$\mu$ is the reduced mass (see the formula and description in
Refs. \cite{Puppe12,Frekers13}).
This means that the values of experimental GT strengths are somehow
model dependent.

In Fig. \ref{48CaGT-}, the calculated running sums of the GT$^-$
strengths [$\Sigma B({\rm GT})$] for $^{48}$Ca are shown as a function
of the excitation energy up to 6.5 MeV, and compared with the data
reported with a red line \cite{Yako09}.
The results obtained with the bare operator (a) are drawn with a blue
dashed line, and those obtained employing the effective GT operator at
LO of the chiral perturbative expansion of ${\mathbf J_A}$ (II) are
plotted with a solid blue line.
The results with the operators (c) and (d), namely with a bare and
effective axial current ${\mathbf J_A}$ expanded up to N3LO, are drawn
with dashed and solid black lines, respectively.

\begin{table}[ht]
  \caption{Experimental \cite{Barabash20,Belli20} and calculated \nmeds~(in
    MeV$^{-1}$) for $^{48}$Ca \dbb~decay.}
\begin{ruledtabular}
\begin{tabular}{ccccccc}
\label{ME_48Ca}
$^{48}$Ca$\rightarrow$$^{48}$Ti& ~ & ~ & ~ & ~ & ~ & ~\\
~& $J^{\pi}_i \rightarrow J^{\pi}_f$ & (a) & (b) & (c) & (d) & Expt \\
\colrule
~& $0^+_1\rightarrow 0^+_1$&0.057&0.048&0.033& 0.019 &$0.042 \pm 0.004$ \\
~& $0^+_1\rightarrow 2^+_1$ & 0.131 & 0.102 & 0.097 & 0.058 & $\leq 0.023$ \\
~& $0^+_1\rightarrow 0^+_2$ & 0.102 & 0.086 & 0.073 & 0.040 & $\leq 2.72$ \\
\end{tabular}
\end{ruledtabular}
\end{table}

It can be seen that the distributions obtained using the (b) and (c)
operators nearly overlap, confirming that the contributions to the
renormalization of the GT operator due to the ChPT expansion of
${\mathbf J_A}$ and to the derivation of a SM effective decay operator
have the same effect in this mass region.

The combination of both effects, which operator (d) accounts for,
results in a good reproduction of the values extracted from data.
It should be pointed out that the theoretical total GT$^-$ strengths
are 24.0, 17.5, 20.9, and 11.2 with operators (a)-(d), respectively,
which should be compared with an experimental one that is
$(15.3\pm2.2)$ and includes a possible contribution from an isovector
spin monopole (IVSM) component \cite{Yako09}.

In Table \ref{ME_48Ca} we report the observed and calculated values of
the \nmeds~ for the \dbb~ decay from $^{48}$Ca ground state into
$^{48}$Ti yrast $J^{\pi}=0^+,2^+$ and yrare $J^{\pi}=0^+$ states, only
for the $0^+_1\rightarrow 0^+_1$ there is a measured value
\cite{Barabash20}, while for the other transitions there are
experimental upper bounds \cite{Belli20}.
We point out that, with respect to the expression in
Eq. (\ref{doublebetameGT}), results are expressed in $g_A^2$ units.

We note that both the experimental and calculated values of \nmed~ are
rather small, compared with those corresponding to the \dbb~ decay of
other nuclides (see Ref. \cite{Barabash20} for a recent review of
current data).
Regarding our calculated \nmeds~ in Table \ref{ME_48Ca}, we have to
underline that their reduced magnitude is mainly related to the fact
that there is some cancellation between the different terms appearing
in the sum in Eq. (\ref{doublebetameGT}).

We may observe that, for the decay between the ground states of
$^{48}$Ca and $^{48}$Ti, the \nmed~ obtained with the bare operator
(a) slightly overestimates the experimental one, and it is three times
larger than the one obtained with the effective operator (d), which
accounts for the ChPT expansion of ${\mathbf J_A}$ as well as the
renormalization due to the configurations outside the chosen $0f1p$
model space.
This corresponds to a quenching factor $q \approx 0.6$, which is
slightly smaller of the empirical value $q=0.7$ commonly considered in
most of calculations that resort to a truncation of the nuclear
degrees of freedom \cite{Martinez-Pinedo96,Suhonen19}.

Regarding the role of the two-body electroweak currents, it is worth
noting that the value of the calculated \nmeds~ with SM effective
operator (d) when compared with the experimental one is very
satisfactory and similar to the result $M^{2\nu}=0.026$ MeV$^{-1}$ we
obtained by deriving \heff~ and \thetaeff from the CD-Bonn potential,
renormalized by way of the \vlwk~ procedure \cite{Coraggio19a}.

\begin{center}
\begin{figure}[ht]
\includegraphics[scale=0.36,angle=0]{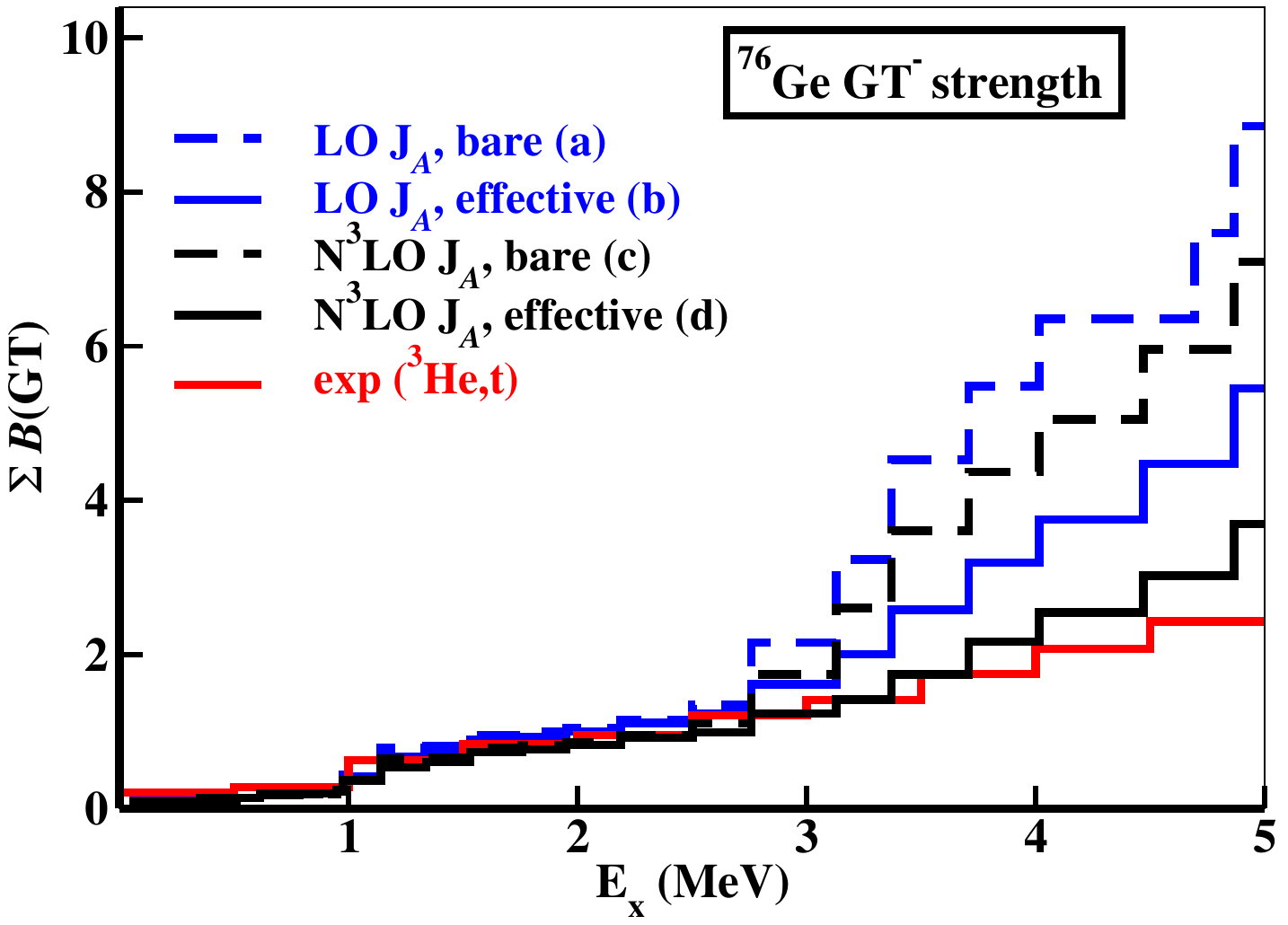}
\caption{Running sums of the $^{76}$Ge $B({\rm GT})$ strengths as a
  function of the excitation energy $E_x$ up to 5 MeV.}
\label{76GeGT-}
\end{figure}
\end{center}

Now, we focus our attention on \zbb-decay candidates which belong to
the $0f_{5/2}1p0g_{9/2}$ shell, namely $^{76}$Ge and $^{82}$Se.

For the calculation of GT-decay properties within such a model space,
we expect that the role of the SM effective operator increases, since
it is well known \cite{Towner87} that spin- and spin-isospin-dependent
operators need larger renormalizations if some of the orbitals
belonging to the model space -- specifically $0f_{5/2}$ and $0g_{9/2}$
-- lack their spin-orbit counterpart.

\begin{center}
\begin{figure}[hb]
\includegraphics[scale=0.36,angle=0]{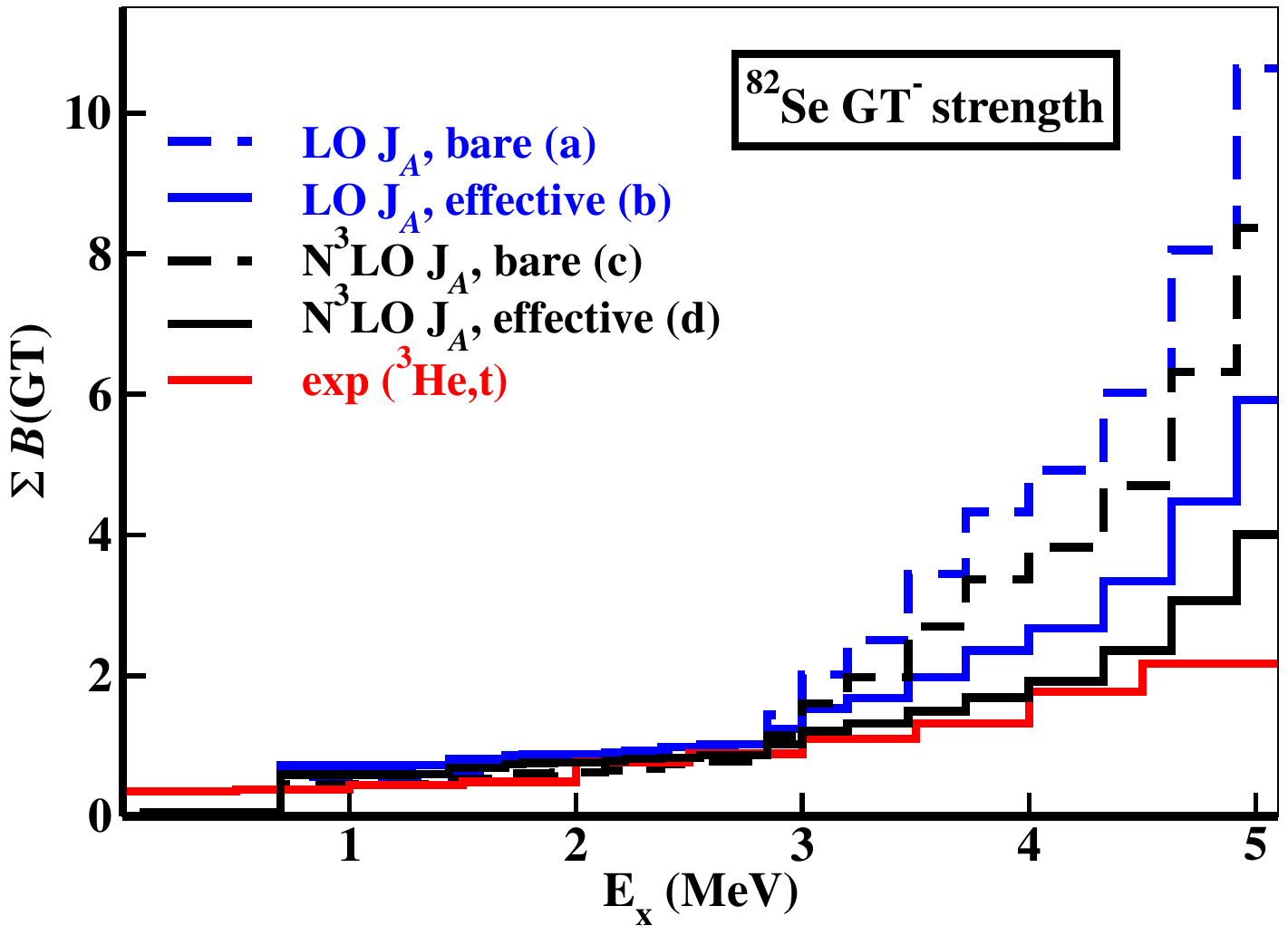}
\caption{Same as in Fig. \ref{76GeGT-}, but for $^{82}$Se.}
\label{82SeGT-}
\end{figure}
\end{center}

We start from the comparison between the experimental
\cite{Thies12a,Frekers16} and the calculated running sums of the
GT$^-$ strengths for $^{76}$Ge and $^{82}$Se, that are reported in
Figs. \ref{76GeGT-} and \ref{82SeGT-}, respectively.
The same labeling as in Fig. \ref{48CaGT-} is used, namely the blue
dashed line represents the calculated values with the bare operator
(a), those obtained with operator (b) are shown with a solid blue
line.
The results with the operators (c) and (d) are plotted with dashed and
solid black lines, respectively.

As can be observed in both Figs. \ref{76GeGT-} and \ref{82SeGT-},
especially for higher energies, only the inclusion of both many-body
renormalizations and higher-order terms in the ChPT expansion can
provide a quite good reproduction of the observed GT$^-$ strength
distribution.
The theoretical total GT$^-$ strengths are 15.8, 10.8, 12.8, and 7.4
with operators (a)–(d), respectively, for $^{76}$Ge and 19.0, 11.4,
14.9, and 7.5 for $^{82}$Se.
At present, there are no available data for these quantities.

\begin{table}[ht]
  \caption{Experimental \cite{Barabash20,Belli20} and calculated \nmeds~(in
    MeV$^{-1}$) for $^{76}$Ge and $^{82}$Se \dbb~decay.}
\begin{ruledtabular}
\begin{tabular}{ccccccc}
\label{ME_76Ge82Se}
$^{76}$Ge$\rightarrow$$^{76}$Se& ~ & ~ & ~ & ~ & ~ & ~\\
~& $J^{\pi}_i \rightarrow J^{\pi}_f$ & (a) & (b) & (c) & (d) & Expt \\
\colrule
~&$0^+_1\rightarrow 0^+_1$&0.211&0.153&0.160& 0.118 &$0.129 \pm 0.004$ \\
~&$0^+_1\rightarrow 2^+_1$& 0.023 & 0.042 & 0.025 & 0.048 & $\leq 0.035$ \\
~&$0^+_1\rightarrow 0^+_2$& 0.009 & 0.086 & 0.016 & 0.063 & $\leq 0.089$ \\
\colrule
 ~ & ~ & ~ & ~ & ~ & ~ & ~ \\
$^{82}$Se$\rightarrow$$^{82}$Kr& ~ & ~ & ~ & ~ & ~ & ~\\
~& $J^{\pi}_i \rightarrow J^{\pi}_f$ & (a) & (b) & (c) & (d) & Expt \\
\colrule
~&$0^+_1\rightarrow 0^+_1$&0.173&0.123&0.136& 0.095 &$0.103 \pm 0.001$ \\
~&$0^+_1\rightarrow 2^+_1$& 0.003 & 0.006 & 0.008 & 0.033 & $\leq 0.020$ \\
~&$0^+_1\rightarrow 0^+_2$& 0.018 & 0.007 & 0.013 & 0.007 & $\leq 0.052$ \\
\end{tabular}
\end{ruledtabular}
\end{table}

Before closing the discussion about the GT-strength distributions, it
is worth emphasizing that, starting from an excitation energy of $\sim
4$ MeV, the experimental distributions are affected by the underlying
contributions arising from the rather structureless tail of the
Gamow-Teller resonance (GTR).
Actually, broad GTRs are observed around $E_x = 11$ MeV and 12.1 MeV
in $^{76}$Ge and $^{82}$Se, respectively, and their contribution to
the total GT strength is not easy to evaluate quantitatively
\cite{Thies12a,Frekers16}.

The central role of two-body electroweak currents shows up also in the
comparison among the calculated \nmeds~ for the \dbb~ decay of the of
the $^{76}$Ge ground state into $^{76}$Se one, as well as for the same
\dbb~decay of $^{82}$Se into $^{82}$Kr, whose values are reported in
Table \ref{ME_76Ge82Se}.

The comparison of experimental $0^+_1\rightarrow 0^+_1$ \nmeds~ with
those calculated by employing SM effective operators (d) is very
satisfactory.
The theoretical values are similar to the ones we have obtained by
deriving \heff~ and \thetaeff~ from the CD-Bonn potential,
renormalized by way of the \vlwk~ procedure \cite{Coraggio19a}.
In this connection, it is worth noting that the renormalization of the
one-body GT operator $\boldsymbol {\sigma \tau}$ due to the use of a
truncated model space is smaller for the \heffs~ employed in the
present paper than the one obtained using the \heffs~ derived in
Ref. \cite{Coraggio19a}.

The calculated values of $0^+_1\rightarrow 2^+_1,0^+_2$ \nmeds~ are
much smaller than the ones involving the transitions between the
ground states, despite being characterized by smaller $Q$ values.
This happens because the contributions corresponding to the different
intermediate states in Eq. (\ref{doublebetameGT}) tend to cancel each
other for such decay branches, while for the decay to the ground
states of $^{76}$Se and $^{82}$Kr they are mostly coherent in sign.

\section{Summary and Outlook}
\label{conclusions}
In this work we have studied for the first time the impact of two-body
electroweak currents, derived with chiral perturbation theory, on the
perturbative renormalization of the shell-model GT-decay effective
operator.

To this end, the electroweak currents have been calculated by way of a
perturbative expansion up to next-to-next-to-next-to-leading order in
chiral perturbation theory, that provides both single- and two-body
components; then the SM effective GT operators have been derived by
way of many-body perturbation theory.
Using such a framework, the SM effective Hamiltonians have been
constructed starting from chiral two and three-body forces -- which
share the same low-energy constants with the expansion of the
electroweak currents -- and then employing the so-called
$\hat{Q}$-box-plus-folded-diagram method to derive the single-particle
energies and two-body matrix elements of the residual Hamiltonian.

This study is a part of a project aiming to calculate reliable nuclear
matrix elements of the \zbb~ decay, therefore we have applied the
present theoretical approach to nuclei of the $0f1p$ and
$0f_{5/2}1p0g_{9/2}$ shells that are of experimental interest for the
detection of such a rare process, namely $^{48}$Ca, $^{76}$Ge, and
$^{82}$Se.

The comparison of our results with experiment for a large set of
observables related to the GT decay -- especially the nuclear matrix
elements of \dbb~ decay -- show that the chiral expansion of the
electroweak currents and the many-body renormalizations, which account
for the configurations outside the model space, share equal merit in
providing a noteworthy reproduction of data.
These results, together with a good reproduction of low-energy
spectroscopic properties of the parent and granddaughter nuclei,
should support the reliability of our future calculation of the
nuclear matrix elements \nmes~ for \zbb~ decay.

In fact the future development of our scientific project is to move
towards a similar study that will involve heavier systems such as
$^{100}$Mo, that are more challenging from the point of view of a
shell-model calculation, as well as the prediction of their \zbb~ with
nuclear matrix elements.
In this way, we could also compare the results based on the ChPT
expansion of nuclear Hamiltonians and electroweak currents with those
we have previously obtained within the realistic shell model starting
from the two-body CD-Bonn potential that was constructed by way of the
meson-exchange theory \cite{Machleidt01b}.

Thus, our following work may also provide important information to
narrow down the spread of the predicted \nmes~ obtained by way of
different approaches.

\begin{acknowledgments}
We thank T. Miyagi for helpful discussion on the inclusion of two-body
currents using the shell-model code KSHELL.
G. D. G. acknowledges the support from the EU-FESR, PON Ricerca e
Innovazione 2014-2020 - DM 1062/2021.
We acknowledge the CINECA award under the ISCRA initiative and under
the INFN-CINECA agreement, for the availability of high performance
computing resources and support.
We acknowledge the support from the National Natural Science
Foundation of China under Grants No. 11835001, No. 11921006, and
No. 12035001.
\end{acknowledgments}

\bibliographystyle{apsrev}
\bibliography{biblio.bib}

\end{document}